\documentclass[a4paper,11pt]{article}
\pdfoutput=1 

\usepackage{jheppub} 
\makeatletter
\gdef\@fpheader{}
\makeatother

\usepackage[T1]{fontenc} 
\usepackage{bm}
\usepackage{amsthm}
\usepackage{xcolor}

\newtheorem{theorem}{Theorem}
\newtheorem{remark}{Remark}

\title{\boldmath Semiparametric Estimation of Physics Parameters and Parton Distribution Functions}

\author[a]{Huanbiao~Zhu,}
\author[b]{Jaco~ter~Hoeve,}
\author[a]{Mikael~Kuusela,}
\author[c]{Maria~Ubiali,}
\author[a]{Larry~Wasserman}

\affiliation[a]{Department of Statistics and Data Science, Carnegie Mellon University\\ 5000 Forbes Avenue, Pittsburgh, USA}
\affiliation[b]{The Higgs Centre for Theoretical Physics, University of Edinburgh,\\[0.1cm]
JCMB, KB, Mayfield Rd, Edinburgh EH9 3FD, Scotland}
\affiliation[c]{DAMTP, University of Cambridge, \\[0.1cm] Wilberforce Road, Cambridge, CB3 0WA, UK}

\emailAdd{huanbiaz@alumni.cmu.edu}
\emailAdd{jaco.ter.hoeve@ed.ac.uk}
\emailAdd{mkuusela@andrew.cmu.edu}
\emailAdd{mu227@cam.ac.uk}
\emailAdd{larry@stat.cmu.edu}

\let\hat\widehat

\abstract{The simultaneous determination of Parton Distribution Functions (PDFs) and fundamental physical parameters, e.~g.\ the strong coupling constant $\alpha_s$, the 
mass of the top quark $m_t$, or Wilson coefficients of an Effective Field Theory (EFT) operators, is essential for the high-precision physics 
programme of the Large Hadron Collider (LHC). 
Yet, simultaneous global fits face significant methodological and computational challenges, rooted in the fundamental tension between estimating a finite-dimensional 
parameter and an infinite-dimensional function. 
In this paper, we address this problem by introducing a semiparametric estimation framework. 
The key advantage of our approach is that it avoids the need to refit the PDFs in the presence of either modified SM parameters or BSM contributions. Starting from standard, 
decoupled estimates of the PDFs and physics parameters, the method computes a bias correction 
\textit{a posteriori} via the efficient influence function, 
yielding a debiased one-step estimator of the finite-dimensional parameter along with uncertainties with valid frequentist coverage. 
This correction rigorously accounts for the interplay between PDF and physics parameters without requiring a computationally 
expensive joint fit. 
As a proof of concept, we apply the method to two toy scenarios preparatory to a realistic global fit: the extraction of 
a dimension-six SMEFT Wilson coefficient from Deep Inelastic Scattering (DIS) pseudo-data based on BCDMS measurements, 
isolating the non-singlet PDF combination $T_3$, and to the determination of the proton total valence $V$ from combined HERA measurements. 
Our results demonstrate that the one-step estimator successfully removes the bias inherent in a preliminary, 
decoupled estimate and achieves confidence intervals with correct nominal coverage, offering a statistically rigorous and computationally 
efficient alternative to full simultaneous PDF fits.}

\usepackage{comment}

\begin{document} 
\maketitle
\flushbottom

\section{Introduction}
\label{sec:intro}

The determination of Parton Distribution Functions (PDFs) relies on the rigorous comparison of proton collider data and their associated uncertainties against 
Standard Model (SM) predictions~\cite{Gao:2017yyd,Kovarik:2019xvh,Amoroso:2022eow,Ubiali:2024pyg,Chiefa:2025loi,Ball:2026dyb}. 
These theoretical predictions, calculated via perturbation theory, depend on fundamental SM parameters and possibly  
on beyond the SM (BSM) parameters.

Conventionally, these parameters are held fixed during a PDF fit, often using values derived from independent, non-hadronic processes such as leptonic production or particle decays. 
However, this decoupled approach is increasingly insufficient for the precision era of the Large Hadron Collider (LHC), especially in light of its high luminosity upgrade (HL-LHC). 
A primary objective of the LHC physics programme is the determination 
of both SM and BSM parameters with high precision. Such a goal demands that PDFs are treated not as static inputs, but rather as dynamic components during the fit. 

To mitigate the risk of double-counting information, a common strategy is to partition experimental data into two distinct sets: one to constrain PDFs at a fixed parameter value, and another to determine parameters using a fixed PDF set. 
Schematically, if $\bm \theta$ are the parameters characterizing the PDFs and $\bm{c}$ is a vector of SM and/or BSM parameters that affect the observable predicted by $T$ in a given model, 
a PDF fit sets the parameters $\bm c$ to some value determined in a separate fit $\bm{c} = {\bm c}^*$ while varying only the PDF parameters $\bm \theta$,
\begin{equation}
    T^{\rm PDF}(\bm \theta, {\bm c}^*) = {\cal L}(\bm \theta) \,\otimes G(\bm c^*),
\end{equation}
where ${\cal L}$ represents the PDF luminosity and $G$ is the forward map including the DGLAP evolution of PDFs and the partonic cross sections, 
which connects the PDFs to the theoretical predictions. 
By contrast, in traditional EFT fits of one or more parameters in $\bm c$, the PDFs are set to some input set $\bm\theta = \bm\theta^*$, which in turn are determined with some $\bm c = \bm c^*$, 
while the parameters $\bm c$ are kept free in the forward map
\begin{equation}
    T^{\rm param}(\bm \theta^*,\bm c) = {\cal L}(\bm \theta^*) \,\otimes G(\bm c).
\end{equation}
This methodology is effective when a process exhibits a strong sensitivity to the parameter of interest but only a weak dependence on the 
underlying PDFs. Classic examples include the Z boson mass $m_Z$, inferred from the position of the dilepton invariant-mass resonance peak in Drell--Yan production, and the Higgs mass, determined from the diphoton invariant-mass peak in $H\to\gamma\gamma$; in both cases the parameter is encoded in a kinematic feature that remains largely invariant to the PDF-dependent overall rate. However, many critical parameters, such as the strong coupling constant $\alpha_s$, defy such a clean separation~\cite{Forte:2020pyp,Stegeman:2025dii}. 
Jet cross-sections, for instance, are highly sensitive to both $\alpha_s$ and the gluon distribution, making them difficult to disentangle. 
In this case reliable unbiased results can only be obtained if $\alpha_s$ and the PDFs are simultaneously determined~\cite{Ball:2025xgq}, 
as opposed to using a fixed PDF set.

A similar challenge arises in the search for new physics. If high-energy datasets, where BSM effects are most likely to manifest, are excluded 
from PDF fits to serve solely as BSM probes, the constraints on large-$x$ gluons and quarks are severely weakened. 
This leaves the community reliant on ageing datasets and the potentially unreliable extrapolation 
of PDF models into high-energy regimes~\cite{Carrazza:2019sec,Greljo:2021kvv,Gao:2022srd,McCullough:2022hzr,CMS:2021yzl,Hammou:2023heg,Kassabov:2023hbm,Hammou:2024xuj,Cole:2026eex}. 
Ideally, a global fit would incorporate all available data to constrain PDFs and physical parameters simultaneously, 
fully accounting for experimental, theoretical, and cross-correlation effects. 
Tools such as {\tt SIMUnet}~\cite{Iranipour:2022iak,Costantini:2024xae}, {\tt Colibri}~\cite{Costantini:2025agd}, the Theory Covariance 
Method~\cite{Ball:2021icz, Ball:2025xgq, Ball:2026qno} and {\tt HERAfitter}~\cite{Shen:2024sci} have started to go in this direction;
{\tt SIMUnet} in particular provides a simultaneous fit of PDFs and an arbitrary number of Wilson coefficients. 
Yet, such an integrated approach presents significant challenges.

Beyond the considerable logistical and computational burdens, a fundamental methodological tension exists: while (B)SM parameters take values in a relatively low-dimensional real space, PDFs are continuous functions whose determination involves complex procedures such as cross-validation and the application of tolerances. 
Combining these distinct statistical frameworks requires a delicate approach to ensure that the precision of parameter extraction is not compromised by 
the methodological ``noise'' inherent in PDF modelling.

Semiparametric inference \cite{hines2022demystifying,kennedy2024semiparametric,nabi2026semiparametric} addresses the estimation of a finite-dimensional parameter of interest in the presence of an infinite-dimensional nuisance parameter. A central object in semiparametric inference is the one-step corrected estimator, which corrects an initial estimate obtained with an approximate nuisance parameter by a single update along the so-called efficient score. This is defined as the component of the score of the parameters of interest that is insensitive to variations of the nuisance parameter, and, as such, removes the first-order bias induced by errors in the estimate of the nuisance. In our setting, the parameter of interest could be SM or BSM parameters, while the PDF $f$ plays the role of the infinite-dimensional nuisance parameter.

In the remainder of this section we focus on the Standard Model Effective Field Theory (SMEFT) framework at dimension-six to parametrise BSM effects~\cite{Brivio:2017vri,Isidori:2023pyp}.
This choice is motivated in part by the fact that the dependence on the Wilson coefficients factorises from the SM cross section to a good approximation, as 
the EFT corrections enter multiplicatively through the partonic cross section while leaving the PDF convolution structure unchanged.
This factorisation, made explicit in Sect.~\ref{sec:model}, is a key property that enables the analytical development of the semiparametric method presented in this work.

Recent work has explored various strategies to disentangle PDF effects in the large-$x$ region from BSM effects parametrised via Wilson coefficients that 
modify the tails of the high-energy experimental distributions. These strategies include conservative fits, or simultaneous fits of 
PDFs and parameters. In some cases, none of these strategies is ideal: conservative fits yield less precise PDFs and simultaneous fits may be too computationally 
burdensome to be included in a simple SMEFT fit, especially as the current tools for simultaneous fits are linked with specific PDF 
parametrizations; NNPDF in the case of {\tt SIMUnet} and HERAPDF in the case of {\tt XFitter}.

We will show that within the semiparametric framework one can estimate the bias to fitting a PDF
nonparametrically.
In some cases, we can also account for the bias due to fitting the PDF with the parameter of interest
set at an incorrect initial value.
This can be done without the need for the complex machinery of a simultaneous fit. A correction term accounting for the use of fixed PDFs is computed 
a posteriori and applied directly to the Wilson coefficient fit. The framework also provides the asymptotic distribution of the debiased parameter estimate, 
which is used to obtain well-calibrated frequentist uncertainties for the Wilson coefficient.

The paper is organized as follows. Sect.~\ref{sec:model} introduces a linear toy model for DIS data in the presence of a simple EFT correction. In Sect.~\ref{sec:methodology}, we introduce and develop the methodology of semiparametric inference in this setting, before applying it as a proof of concept to extract an EFT parameter from BCDMS pseudo-data in Sect.~\ref{sec:application_BCDMS} and HERA pseudo-data in Sect.~\ref{sec:application_HERA}. We summarise and conclude in Sect.~\ref{sec:conclusion}.

\section{PDF Model}
\label{sec:model}
We start by introducing the forward map that we adopt throughout this work to map from the space of PDFs to the space of observables. 
We focus on Deep Inelastic Scattering (DIS) data in which the PDFs enter linearly, thus allowing a fully analytic treatment of the 
semiparametric method. Possible generalisations to the proton-proton case are deferred to our conclusions. Furthermore, we identify 
the parameter of interest as a Wilson coefficient associated with a specific energy-growing operator in the dimension-six SMEFT 
expansion. This choice has the advantage that the parameter dependence is entirely contained in the partonic cross section and that, to a good approximation,  
the dependence on the parameter is encapsulated in a multiplicative factor.

\subsection{Forward Map}
\label{subsec:forward-map}
The theoretical prediction $T^J$ for a DIS  observable $J$ at kinematic point $(x, Q^2)$ is obtained by convolving\footnote{The convolution operator in Eq.~\eqref{eq:forward_map} is defined as
\begin{equation}
    (f \otimes h)(x) \equiv \int_x^1\frac{dy}{y}f(y)h\left(\frac{x}{y}\right) \, .
    \label{eq:convolution}
\end{equation}} a known response kernel $\sigma^J$ with the PDF $f$,
\begin{align}
    \nonumber T^J(x, Q^2; f, c) &= \sum_{i=1}^{n_{\rm parton}}\sigma_i^{J}(x, Q^2, c)\otimes f_i(x, Q^2)\\
    &= \sum_{i,j=1}^{n_{\rm parton}}\sigma_i^J(x, Q^2, c) \otimes \Gamma_{ij}(Q^2, Q_0^2) \otimes f_j(x, Q_0^2) \, ,
    \label{eq:forward_map}
\end{align}
where the sum runs over the $n_{\rm parton}$ active flavours at the scale $Q$ and $\Gamma_{ij}(Q^2, Q_0^2)$ denotes the evolution kernel operator that evolves the PDF $f$ from the parameterisation scale $Q_0$ to the hard scattering scale $Q$ as determined by the DGLAP evolution equations. The response kernel $\sigma_i^J$ can be computed in perturbation theory and may furthermore depend on an EFT parameter of interest $c\in\mathbb{R}$ that we assume to be real by restricting ourselves to CP even dimension-6 operators.

To speed up the computation of the integral in Eq.~\eqref{eq:forward_map}, one can write $f_i(x,Q^2)$ on a given grid in $x$ and $Q^2$, namely $\{x_\alpha, Q^2_\beta\}$ 
and write it as a sum of interpolating functions times the function itself computed at the grid points, 
\begin{equation}
    f_i(x,Q^2) = \sum_{\alpha=1}^{n_{\rm grid}}\sum_{\beta=1}^{n_{\rm grid}} f_{i, \alpha\beta}\,{\cal I}_\alpha(x){\cal I}_\beta(Q^2),
\end{equation}
which can be expressed with respect to the parameterisation scale by interpolating the DGLAP evolution kernel,
\begin{equation}
    f_{i, \alpha\beta} = \sum_{j=1}^{n_{\rm parton}}\sum_{\rho=1}^{n_{\rm grid}} \Gamma_{ij, \alpha\beta\rho} f_j(x_{\rho}, Q_0^2)\, .
\end{equation}
Doing so, the integral in Eq.~\eqref{eq:forward_map} can be written as
\begin{equation}
\label{eq:FKtab}
    T^{J}(x, Q^2; f, c) = \sum_{i, j=1}^{n_{\rm parton}}\sum_{\alpha, \beta, \rho=1}^{n_{\rm grid}}\left(\int_x^1 \frac{dy}{y} \sigma_i^J\left(\frac{x}{y}, Q^2, c\right) \Gamma_{ij, \alpha\beta\rho} {\cal I}_\alpha(y) {\cal I}_\beta(Q^2)\right)f_{j}(x_\rho, Q_0^2) \, .
\end{equation}
The integral in the parentheses can be pre-computed before doing a fit and is called Fast-Kernel (FK) table in NNPDF jargon \cite{Ball:2010de}. The forward model then takes the form
\begin{equation}
\label{eq:FKsum}
    T^{J}(x, Q^2; f, c) = \sum_{i=1}^{n_{\rm parton}}\sum_{\alpha=1}^{n_{\rm grid}}\,  ({\rm FK})^J_{i\alpha}(x, x_\alpha, Q^2, Q_0^2, c)f_i(x_\alpha, Q_0^2) \, , 
\end{equation}
where the computation of the integral at each iteration of the fitted functions $f_i$ is substituted by a precomputed finite sum over the collection of grid points. Additionally, for the purpose of this work, we are only working with a single PDF and single observable, so we drop the indices $i$ and $J$ in the following sections for simplicity. 

\subsection{EFT Expansion of the Response Kernel}
\label{subsec:eft_toy_example}
The response kernel $\sigma$ admits the following expansion up to quadratic order in the presence of a single dimension-six EFT parameter $c$,
\begin{equation}
\label{eq:BSMdecomp}
    \sigma\!\left(x, Q^2, c\right)
    = \sigma_{\rm SM}\!\left(x, Q^2\right)
      \Bigl(1 + c\,k_{\rm lin}(Q^2) + c^2\,k_{\rm quad}(Q^2)\Bigr),
\end{equation}
where we have assumed for simplicity that the functions $k_{\rm lin}$ and $k_{\rm quad}$ are independent of $x$ and thus depend only on $Q^2$. 
A key feature of the decomposition in Eq.~\eqref{eq:BSMdecomp} is that the correction functions $k_{\rm lin}$ and $k_{\rm quad}$ are treated as 
independent of the PDFs. This is motivated by their interpretation as ratios of EFT to SM partonic cross sections: since they are defined at the 
level of hard-scattering coefficients, they carry no PDF dependence at leading order. At higher orders, evaluating such ratios numerically requires 
a reference PDF set, but the PDF dependence largely cancels between numerator and denominator. A detailed study of this cancellation in the context 
of high-energy Drell-Yan tails, presented in Appendix C of Ref.~\cite{Greljo:2021kvv}, found the residual PDF sensitivity to be at the permil to percent level, 
justifying the approximation adopted here.

\begin{figure}[htbp]
    \centering
    \includegraphics[width=0.9\textwidth]{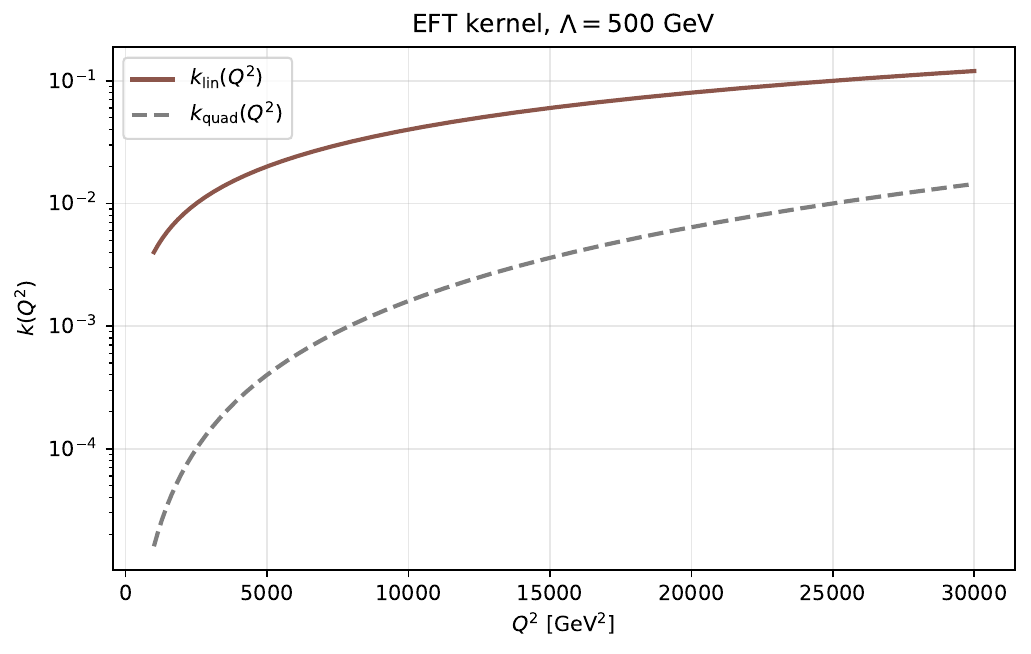}
    \caption{The EFT coefficient functions $k_{\rm lin}$ and $k_{\rm quad}$ defined in Eq.~\eqref{eq:eft_corrections} as a function of $Q^2$.}
    \label{fig:EFT_kernel}
\end{figure}

As a simple toy model that permits energy-growing effects characteristic for a four-fermion operator in the SMEFT 
(see Ref.~\cite{Bissolotti:2023vdw} for a more realistic study of four-fermion operators on DIS observables), we take
\begin{equation}
    k_{\rm lin}(Q^2) = \frac{Q^2}{\Lambda^2}, \qquad k_{\rm quad}(Q^2) = \frac{Q^4}{\Lambda^4},
    \label{eq:eft_corrections}
\end{equation}
where $\Lambda$ denotes the order of the scale of new physics.
Since we will be analysing DIS data in which the highest $Q^2$ value is attained at $Q^2=30000$ GeV$^2$, in our toy model 
we set $\Lambda = 500$ GeV such that the scale of potential new physics is well above the highest $Q^2$ bin. The dependence of the EFT coefficient functions $k_{\rm lin}$ and $k_{\rm quad}$ on $Q^2$ for this choice of $\Lambda$ is illustrated in Fig.~\ref{fig:EFT_kernel}. Note, however, 
that one has freedom in defining the values of both $c$ and $\Lambda$ as they always appear as the fixed combination 
$c/\Lambda^2$. In the applications considered in Sect.~\ref{sec:application_BCDMS} and Sect.~\ref{sec:application_HERA}, 
the experiments probe different ranges of $Q^2$. When generating BSM pseudo data, we therefore choose $c$ separately for each application so that the induced 
BSM effect is sufficiently pronounced over the relevant kinematic range to clearly illustrate its impact on the inference results.

Substituting \eqref{eq:BSMdecomp} into \eqref{eq:forward_map}, the forward model factorises as
\begin{equation}
\label{eq:factorization}
    T(x,Q^2;f,c) = \nu(c, Q^2)T_{\rm SM}(x,Q^2;f),
\end{equation}
where
\begin{equation}
    \nu(c,Q^2)=1+k_{\rm lin}(Q^2)c+k_{\rm quad}(Q^2)c^2, \quad T_{\rm SM}(x,Q^2;f)\equiv
    T(x, Q^2; f, 0)\, .
\end{equation}
Note that the functions $k_{\rm lin}$ and $k_{\rm quad}$ entering the response kernel $\sigma$ may in general depend explicitly on both $x$ and $Q^2$. The bias-correction procedure in this work is based on the simplified model in Eq.~\eqref{eq:factorization}. Nevertheless, the semiparametric estimation procedure described in the next section can also be applied to this more general setting. The main difference is that the corresponding nuisance tangent space and efficient influence function must be rederived under the more general model.

In the following, it will be useful to consider pseudo-data generated according to a known underlying law. Given $n$ measurements with kinematic points $\{(x_i, Q_i^2)\}_{i=1}^n$ and an experimental covariance matrix $\bm{\Sigma} \in \mathbb{R}^{n \times n}$, we denote the corresponding theory predictions in the SM by
\begin{equation}
\mathbf{T}_{\rm SM}(f) = [T_{\rm SM}(x_1, Q_1^2;f), \dots, T_{\rm SM}(x_n, Q_n^2;f)]^\top \in \mathbb{R}^n, \,
\end{equation}
which, following Eq.~\eqref{eq:factorization}, receives an overall multiplicative EFT correction as
\begin{equation}
 \mathbf{T}(c,f) = \bm{\nu}(c) \odot \mathbf{T}_{\rm SM}(f),
\end{equation}
where $\odot$ denotes the elementwise product and we have defined the vector of EFT corrections as
\begin{equation}
\bm{\nu}(c) \equiv [\nu(c, Q_1^2), \dots, \nu(c, Q_n^2)]^\top \in \mathbb{R}^n.    
\end{equation}
We then generate pseudo-data $\bm{y}$ normally distributed around $\mathbf{T}(c,f)$ with covariance $\bm{\Sigma}$,
\begin{equation}
\label{eq:data_model}
    \bm{y} \sim \mathcal{N}\!\bigl(\mathbf{T}(c,f),\, \bm{\Sigma}\bigr),
\end{equation}
which we will refer to as ``data'' from now on.  

\section{Semiparametric Inference in the SMEFT}
\label{sec:methodology}
To estimate the Wilson coefficient $c$, we make use of a statistical approach known as {\em semiparametric inference} \cite{vandervaart1998, tsiatis2006}, which is a general framework for estimating a finite-dimensional parameter in the presence of an infinite-dimensional nuisance parameter. In our setting, the finite-dimensional parameter of interest is the Wilson coefficient $c$, while the PDF $f$ plays the role of an infinite-dimensional nuisance parameter. The procedure requires a preliminary, possibly biased, estimate of $f$ and $c$. The preliminary estimate of $f$ could come from a standard PDF fit such as the CT global fit~\cite{Hou:2019}, MSHT20~\cite{Bailey:2020ooq} or NNPDF4.0~\cite{ball2022}.  We then use the efficient influence function defined subsequently to correct the bias in the preliminary estimate of $c$. Specifically, the procedure is as follows:
\begin{enumerate}
    \item Obtain preliminary estimate $\hat f_0$ by fitting the data $\bm{y}$ with a standard PDF fit.
    \item Obtain preliminary estimate $\hat c_0$ based on the fitted $\hat f_0$ and the data.
    \item Compute the efficient influence function $\varphi(\bm{y};{c}_0,{f}_0)$ for $c$.
    \item Construct the debiased estimator $\hat c_{\rm one-step} = \hat c_0 + \varphi(\bm{y}; \hat c_0, \hat f_0)$. 
\end{enumerate}
A key benefit of this approach is that it avoids the need to refit the PDFs in the presence of the BSM contribution: the bias correction is applied once, {\it a posteriori}, to the preliminary estimate. The resulting estimator $\hat c_{\rm one-step}$ is often called the ``one-step estimator'' since it takes one correction step in the direction suggested by the efficient influence function. Moreover, as we will show, this procedure also yields confidence intervals for $c$ with valid frequentist coverage under suitable regularity conditions.

We begin by giving a brief introduction to semiparametric inference and then describe how to apply the method to our problem setting.

\medskip

\subsection{Semiparametric Inference}
\label{sec:semiparametric_intro}
The goal of semiparametric inference
\cite{vandervaart1998, tsiatis2006}
is to estimate a finite-dimensional parameter $c$
when there is an infinite-dimensional nuisance parameter $f$.
For example, suppose that
$Y_1,\ldots, Y_n$ are an i.i.d.\ sample from a density $f(y)$
and that we want to estimate
\begin{equation}
c = \psi(f)=\int f^2(y)dy. 
\end{equation}
The parameter $c$ is scalar but
the unknown density $f$ is infinite-dimensional. A natural first choice is to estimate $f$ by some non-parametric estimator $\hat f$, such as a kernel density estimator, and then form the {\em plugin estimator}
\begin{equation}
\hat c_{0} = \int \skew{4}{\hat}{f}^{\,2}(y) dy \, .
\end{equation}
This estimator is
very biased and is far from optimal. Semiparametric theory improves on this by adding an influence function correction. The resulting one-step estimator takes the form
\begin{equation}
\label{eq:one_step_iid}
\hat c_{\rm one-step} = \hat c_{0} + \frac{1}{n}\sum_{i=1}^n \varphi\left(Y_i,\skew{4}{\hat}{f}\right),
\end{equation}
where
$\varphi$ is the {\em efficient influence function}, which we define below.
This estimator has the property that
$\sqrt{n}\,(\hat c_{\text{one-step}} - c)$
converges to a Normal distribution with zero mean and variance $\sigma^2 = \mathbb{E}[\varphi^2(Y,f)]$. This estimator is optimal in the sense that any other well-behaved estimator $\tilde c$ will have an asymptotic variance that is larger than $\sigma^2$.

\begin{figure}
    \centering
    \includegraphics[width=0.7\linewidth]{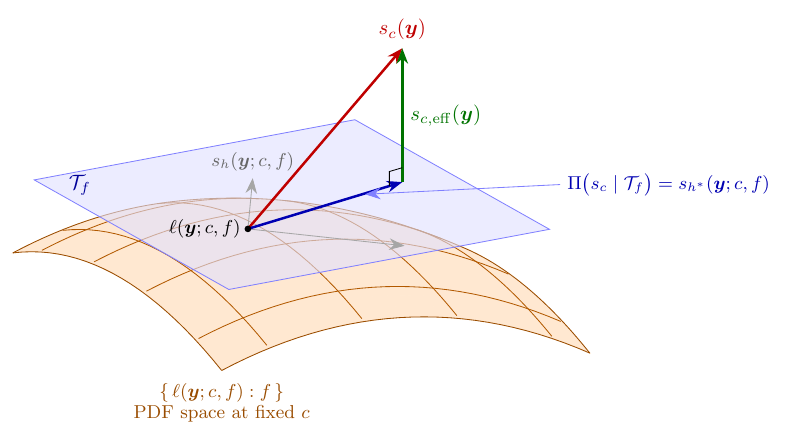}
    \caption{Geometric representation of the construction of the efficient score $s_{c, {\rm eff}}$ for a log-likelihood $\ell({\bm y}; c,f)$. The curved surface is the PDF space at fixed $c$, obtained by varying $f$, and its nuisance tangent space $\mathcal{T}_f$ is the span of all nuisance scores $s_h$. The score $s_c$ decomposes into a component along $\mathcal{T}_f$, the projection $\Pi(s_c|\mathcal{T}_f)$, which can be mimicked by locally varying the PDFs, and a component orthogonal to $\mathcal{T}_f$, the efficient score $s_{c, {\rm eff}}$ that is insensitive to local variations in the direction of the nuisance function $f$.}
    \label{fig:construct_sc_eff}
\end{figure}

Fig.~\ref{fig:construct_sc_eff} illustrates how to construct the efficient influence function by means of the efficient score. Typically, one forms one-dimensional subfamilies $f_\epsilon$ passing through the true distribution at $\epsilon=0$. For example, one may write 
\begin{equation}
f_\epsilon = f(1+\epsilon h)\, , 
\end{equation}
where $h\in\mathcal{H}$ is a perturbation in the space of allowed perturbative corrections $\mathcal{H}$ so that $f_\epsilon$ remains a valid density.\footnote{We refer to Ref.~\cite{kennedy2024semiparametric} for the conditions $h$ must satisfy for $f_\epsilon$ to remain a valid density.}
Each choice of function $h$ defines a different subfamily.
In some problems, including our SMEFT example from Sect.~\ref{subsec:eft_toy_example}, the finite-dimensional parameter of interest and the infinite-dimensional nuisance parameter enter the statistical model independently.
This means that the log-likelihood $\ell$ can be written explicitly as
\begin{equation}
\ell(y;c,f)=\log p(y;c,f).
\end{equation}
Taking the derivative of $\ell$ with respect to $c$ gives the \textit{score} for the parameter of interest 
\begin{equation}
s_c(y;c,f)=\frac{\partial}{\partial c}\ell(y;c,f).
\end{equation}
For each parametric submodel $f_\epsilon$ through $f$, we also obtain a nuisance score
\begin{equation}
s_h(y;c,f)=\frac{\partial}{\partial \epsilon}\log p(y;c,f_\epsilon)\Biggm|_{\epsilon=0}.
\end{equation}
The closure of the span of all such nuisance scores over $h$ is called the nuisance tangent space and is denoted by $\mathcal T_f$. The \textit{efficient score} for $c$ is the component of $s_c$ orthogonal to the nuisance tangent space:
\begin{equation}
s_{c,\text{eff}}(y;c,f)=s_c(y;c,f) - \Pi(s_c(y;c,f) \mid \mathcal{T}_f)
\end{equation}
where $\Pi(s_c|\mathcal T_f)$ denotes the $L^2(P)$ projection of $s_c$ onto $\mathcal T_f$. The efficient influence function $\varphi$ is then given by
\begin{equation}
\varphi(y;c,f)=\frac{s_{c,\text{eff}}(y;c,f)}{\mathbb{E}[s_{c,\text{eff}}^2(Y;c,f)]} \, ,
\end{equation}
where the expectation is defined with respect to the density $f$.
For a more comprehensive review of semiparametric inference and efficient influence functions, see, for example, \cite{hines2022demystifying,kennedy2024semiparametric,nabi2026semiparametric}.

\subsection{Estimation Strategy}
\label{sec:estimation_strategy}
For our application, we consider a general PDF function space,
\begin{equation}
    \mathcal{F}=L^2([0,1]),
\end{equation}
and optionally impose the normalization constraint
\begin{equation}
    \int_0^1 f(y)\,dy=\kappa,
\end{equation}
where $\kappa$ is a positive real constant. The constrained formulation allows known PDF sum rules to be incorporated directly, 
such as the valence sum rule for $V,V_3,V_8$, while the unconstrained formulation allows to apply the methodology to the triplet 
PDFs like $T_3$.
Based on our model given by \eqref{eq:data_model} with the corresponding space
\begin{equation}
{\cal M} = \Bigl\{ \mathbf{T}(c,f):\ c\in\mathbb{R}, \ f\in {\cal F}\Bigr\},
\end{equation}
our goal is to estimate $c$ and $f$ from the data vector $\bm{y}$. One of the most common approaches is through coupled maximum likelihood estimation (MLE). That is, we seek $(\hat c_0, \hat f_0)$
maximizing the log-likelihood (up to a constant)
\begin{equation}
\ell(c,f)=-\frac{1}{2}(\bm{y} -  \mathbf{T}(c,f))^\top
\bm{\Sigma}^{-1}(\bm{y} -  \mathbf{T}(c,f))
\end{equation}
over all $c$ and all $f\in {\cal F}$.
Because ${\cal M}$ is a large, nonparametric class,
$\hat c_0$ may have first-order bias induced by estimation of $f$.
The debiased one-step estimate of $c$ is
\begin{equation}
    \hat{c}_{\text{one-step}} = \hat{c}_0 + \varphi(\bm{y};\, \hat{c}_0,\, \hat{f}_0),
    \label{eq:one-step-estimate}
\end{equation}
where
$\varphi$ is the efficient influence function for $c$, which we will derive in Sect.~\ref{sec:EIF}. Unlike the usual i.i.d.\ formulation reviewed in Sect.~\ref{sec:semiparametric_intro}, our observation is the full data vector $\bm{y}$, whose entries may not be i.i.d. due to experimental and theoretical correlations. This means that the log-likelihood cannot be written as a single sum over individual observations (events).  Therefore, we compute the one-step correction from the efficient influence function of the full sample $\bm y$ instead of as an empirical average over individual observations in Eq.~\eqref{eq:one_step_iid}.

Under some regularity conditions \cite{kennedy2024semiparametric}, the one-step estimator is asymptotically efficient. In particular, we have
\begin{equation}
\sqrt n(\hat c_{\rm one-step}-c)
\rightsquigarrow
N\left(0,\sigma^2\right)
\end{equation}
as $n\rightarrow\infty$, where $n$ is the total number of elements in $\bm{y}$. The variance $\sigma^2$ is the semiparametric efficiency bound for the scaled estimator, which is a non-degenerate limiting constant and does not depend on $n$. Equivalently, under the same normalization, the full-sample influence-function satisfies
\begin{equation}
\label{eq:var_EIF}
n\mathbb{E}[\varphi^2(\bm{y};c,f)]\to \sigma^2.
\end{equation} 
The factor $1/n$ appears because $\varphi(\bm{y};c,f)$ is the influence-function correction for the full data vector $\bm{y}$. This is analogous to the fact that, in the i.i.d. setting from Eq.~\eqref{eq:one_step_iid}, the variance of the empirical average of per-observation influence functions is the variance of a single influence function divided by $n$. Therefore, a $1-\alpha$ confidence interval for $c$ can be constructed as 
\begin{equation}
    \hat{c}_{\rm one-step} \pm \frac{z_{\alpha/2} \hat\sigma}{\sqrt{n}},
    \label{eq:ci_one_step}
\end{equation}
where $\hat{\sigma}=\sqrt{n\mathbb{E}[\varphi^2(\bm{y};\hat c_0,\hat f_0)]}$ and $z_{\alpha/2}$ is the $(1-\alpha/2)$-quantile of the standard normal distribution.

Regarding the preliminary estimators $\hat{c}_0$ and $\hat{f}_0$, 
ideally
these are obtained
to maximize $\ell(c,f)$ simultaneously over $f$ and $c$. 
Since $f$ is being treated nonparametrically,
the resulting estimator $\hat c_0$ will be biased
and our procedure will reduce this bias.
However, as discussed in the introduction, this joint optimization can be challenging in practice. In this work, we focus on the more realistic decoupled fitting procedure, where we minimize over $c$ with $f$ fixed at a preliminary estimated $\hat f_0$ based on the Standard Model. Specifically, we find
\begin{equation}
    \hat{f}_0=\arg\max_f\ell(c=0,f), \quad \hat{c}_0=\arg\max_c\ell(c,\hat{f}_0).
\end{equation}
In practice, this introduces some extra bias to $\hat c_0$.
As long as this extra bias is not too large,
our procedure will still remove this bias. In Appendix~\ref{sec:one-step-validity}, we establish the asymptotic validity of the one-step estimator under regularity conditions that do not explicitly require the preliminary estimates to be obtained from the joint MLE. Furthermore, our applications from Sect.~\ref{sec:application_BCDMS} and \ref{sec:application_HERA} demonstrate that the proposed estimator performs well in practice with decoupled MLE.

\subsection{Efficient Influence Function}
\label{sec:EIF}
As shown in Sect.~\ref{sec:estimation_strategy}, the efficient influence function $\varphi$ is the key to constructing the debiased estimator of $c$. In this section, we derive the efficient influence function $\varphi$. For notational convenience, define the linear operators
\begin{equation}
    K_ch := \bm{\nu}(c) \odot \mathbf{T}_{\rm SM}(h), \qquad
    \dot{K}_ch := \bm{\nu}'(c) \odot \mathbf{T}_{\rm SM}(h),
    \label{eq:linear_ops_Kch}
\end{equation}
where $\mathbf{T}_{\rm SM}(h)$ denotes the vector of SM predictions evaluated at the kinematic points $\{(x_i, Q_i^2)\}_{i=1}^n$ with input function $h$. With this notation, the mean vector is $\mathbf{T}(c,f)=K_c f$.

The first step to derive the efficient influence function is to compute the score functions for $c$ and $f$. The score function for $c$ is given by the partial derivative of the log-likelihood with respect to $c$:
\begin{equation}
    s_c(\bm{y}) = \frac{\partial}{\partial c}\ell(c, f)
    = (\dot{K}_cf)^\top \bm{\Sigma}^{-1}(\bm{y} - K_cf).
\end{equation}
Similarly, the score for $f$ in a perturbation direction $h$ is obtained via the parametric submodel $f_\epsilon = f + \epsilon h$:
\begin{equation}
    s_f(\bm{y}, h)
    = \frac{\partial}{\partial\epsilon}\ell(c,f_\epsilon)\Big|_{\epsilon=0}
    = (K_ch)^\top {\bm\Sigma}^{-1}(\bm{y} - K_cf).
\end{equation}
The efficient score for $c$ is then given by the projection of $s_c$ onto the orthogonal complement of the nuisance tangent space $\mathcal{T}_f$ (i.e., the space spanned by the score for $f$), which is defined with respect to the $L_2(P)$ inner product $\langle g_1,g_2\rangle = \mathbb{E}[g_1(\bm{y})g_2(\bm{y})]$. Thus,
\begin{equation}
    s_{c,\text{eff}}(\bm{y})
    = s_c(\bm{y}) - \Pi(s_c(\bm{y}) \mid \mathcal{T}_f)
    = s_c(\bm{y}) - s_f(\bm{y}, h^*),
\end{equation}
where $h^*$ is the perturbation of $f$ whose score is closest to the score $s_c$, i.e., the variation of $f$ that best resembles the effect of a shift in $c$ on the likelihood, 
\begin{equation}
    \mathcal{T}_f = \{s_f(\bm{y}, h):h\in\mathcal{H}\}
    ,\;\; h^* = \operatorname*{arg\,min}_{h\in\mathcal{H}}
    \mathbb{E}\bigl[(s_c - s_f(h))^2\bigr].
\end{equation}
Since
\begin{equation}
    \mathbb{E}[(s_c(\bm{y})-s_f(\bm{y}, h))^2] = (\dot{K}_cf-K_ch)^\top\bm{\Sigma}^{-1}(\dot{K}_cf-K_ch),
\end{equation}
the projection problem is equivalent to minimizing
\begin{equation}
\label{eq:projection-objective}
    (\dot{K}_c f-K_c h)^\top
    \bm{\Sigma}^{-1}
    (\dot{K}_c f-K_c h)
\end{equation}
over the admissible perturbation directions $h$. The set of admissible directions depends on whether the normalization constraint is imposed. In the unconstrained model, $h$ may be any element of the nuisance function space. The minimizer is therefore
\begin{equation}
\label{eq:hstar-unconstrained}
    h^*
    =
    \bigl(K_c^*\Sigma^{-1}K_c\bigr)^{-1}
    K_c^*\bm{\Sigma}^{-1}\dot{K}_c f,
\end{equation}
where $K_c^*$ denotes the adjoint of $K_c$. In the constrained model, the normalization condition
\begin{equation}
    \int_0^1 f(y)\,dy=\kappa
\end{equation}
must also hold along the parametric submodel
$f_\epsilon=f+\epsilon h$. Consequently, admissible perturbations must satisfy
\begin{equation}
    \int_0^1 h(y)\,dy=0.
\end{equation}
The constrained projection problem is thus
\begin{equation}
\label{eq:constrained-projection}
    h^*
    =
    \operatorname*{arg\,min}_{h:\,\int_0^1 h(y)\,dy=0}
    (\dot{K}_c f-K_c h)^\top
    \bm{\Sigma}^{-1}
    (\dot{K}_c f-K_c h).
\end{equation}
Introducing a Lagrange multiplier $\lambda$, we define 
\begin{equation}
    \mathcal{L}(h,\lambda) = (\dot{K}_cf-K_ch)^\top{\bm\Sigma}^{-1}(\dot{K}_cf-K_ch) + 2\lambda \int h(y)dy.
\end{equation}
Having the Gâteaux derivative of $\mathcal{L}$ with respect to $h$ to vanish in every direction $g$ yields the first-order condition
\begin{equation}
K_c^*{\bm\Sigma}^{-1}K_c h^*
=
K_c^*{\bm\Sigma}^{-1}\dot{K}_c f
-
\lambda\mathbf{1},
\end{equation}
where $\mathbf{1}$ denotes the constant function equal to one. Assuming that $K_c^*{\bm\Sigma}^{-1}K_c$ is invertible, the solution is
\begin{equation}
\label{eq:h*}
    h^* = (K^*_c\bm{\Sigma}^{-1}K_c)^{-1}K^*_c\bm{\Sigma}^{-1}\dot{K}_cf - \lambda(K^*_c\bm{\Sigma}^{-1}K_c)^{-1}\mathbf{1},
\end{equation}
 The Lagrange multiplier $\lambda$ can be found as
\begin{equation}
    \lambda = \frac{\langle\mathbf{1},(K^*_c\bm{\Sigma}^{-1}K_c)^{-1}K^*_c\bm{\Sigma}^{-1}\dot{K}_cf\rangle}{\langle\mathbf{1},(K^*_c\bm{\Sigma}^{-1}K_c)^{-1}\mathbf{1}\rangle}
\end{equation}
where the inner product $\langle\cdot, \cdot\rangle$ is defined as $\langle g, h\rangle = \int g(y)h(y)dy$.
In both the constrained and unconstrained case, substituting $h^*$ into the expression for the efficient score gives
\begin{equation}
\label{eq:efficient_score}
    s_{c,\text{eff}}(\bm{y}) = (\dot{K}_cf - K_ch^*)^\top {\bm\Sigma}^{-1}(\bm{y} - K_cf),
\end{equation}
and the efficient influence function (EIF) is then given by
\begin{equation}
\label{eq:EIF}
    \varphi(\bm{y};c,f) = \frac{s_{c,\text{eff}}(\bm{y})}{\mathbb{E}[s_{c,\text{eff}}^2(\bm{y})]}
    = \frac{(\dot{K}_cf - K_ch^*)^\top \bm{\Sigma}^{-1}(\bm{y} - K_cf)}
           {(\dot{K}_cf - K_ch^*)^\top \bm{\Sigma}^{-1}(\dot{K}_cf - K_ch^*)}.
\end{equation}
The one-step estimate of $c$ then follows by substituting 
Eq.~\eqref{eq:EIF} into Eq.~\eqref{eq:one-step-estimate}.

\subsection{Basis Expansion}
In practice, existing PDFs have been fitted to SM predictions. In our application, to introduce BSM effects, we treat the NNPDF4.0 PDF set \cite{ball2022} as the ground truth, and generate pseudo-data based on this PDF set and a specified Wilson coefficient $c\neq0$. Then, as a first step in the estimation procedure, we will need to obtain some preliminary estimates of $f$ and $c$ from the pseudo-data. Fitting the PDF $f$ directly at the grid points $x_\alpha$ can be an ill-posed problem whenever the forward map $({\rm FK})$ is not full rank, in which case some combination of grid values $f(x_\alpha)$ lies in the null space of the forward map and is therefore invisible to the data. In such cases, the inverse problem needs to be regularised by introducing, for example, $m$ B-spline basis functions \cite{deboor1978, hastie2009} to represent the PDF $f$. Specifically, in this work, we fit a single PDF flavor $f$ by expanding $yf(y)$ in terms of $m$ basis functions $\{B_k\}_{k=1}^m$ on $[0,1]$ at the scale $Q_0$,
\begin{equation}
    yf(y) = \sum_{k=1}^m \beta_k B_k(y), \qquad \bm{\beta} = [\beta_1, \dots, \beta_m]^\top \in \mathbb{R}^m \, ,
    \label{eq:pdf_spline_expansion}
\end{equation}
and define the design matrix $A \in \mathbb{R}^{n \times m}$ by 
\begin{equation}
    A_{ik} = \sum_{\alpha=1}^{n_{\rm grid}}
    ({\rm FK})_{\alpha}(x_i, x_\alpha, Q_i^2, Q_0^2)\, \frac{B_k(x_\alpha)}{x_\alpha} \, ,
    \label{eq:design_matrix_A}
\end{equation}
where the number of basis functions $m$ is such that the design matrix $A$ is full rank. The SM prediction can then be written as 
\begin{equation}
    \mathbf{T}_{\rm SM}(f) = A\bm{\beta} \, .
\end{equation}
Similarly, the operators $K_c$ and $\dot{K}_c$ of Eq.~\eqref{eq:linear_ops_Kch} become matrices acting on $\bm{\beta}$, 
\begin{equation}
    K_c\bm{\beta} = \bm{\nu}(c) \odot (A\bm{\beta}),
    \qquad
    \dot{K}_c\bm{\beta} = \bm{\nu}'(c) \odot (A\bm{\beta}) \, ,
\end{equation}
while the log-likelihood reduces to
\begin{equation}
    \ell(c,\bm{\beta})
    = -\frac{1}{2}\bigl(\bm{y} - K_c\bm{\beta}\bigr)^\top \bm{\Sigma}^{-1}\bigl(\bm{y} - K_c\bm{\beta}\bigr) + \mathrm{const}.
\end{equation}
If $f$ is required to satisfy the normalization condition
$\int_0^1 f(y)dy=\kappa$, then the coefficient vector $\bm{\beta}$ must satisfy
\begin{equation}
\bm{w}^\top\bm{\beta}=\kappa,
\qquad
w_k=\int_0^1 \frac{B_k(y)}{y}dy.
\end{equation}
In this finite-dimensional representation, a nuisance
perturbation $h(y)$ is represented by its spline coefficient
vector $\bm{\gamma}\in\mathbb{R}^m$, such that
\begin{equation}
    yh(y)=\sum_{k=1}^m \gamma_k B_k(y),
    \qquad
    \bm{\gamma}
    = [\gamma_1,\dots,\gamma_m]^\top.
\end{equation}
The scores for $c$ and $\bm{\beta}$ are
\begin{align}
    s_c(\bm{y}) &= (\dot{K}_c\bm{\beta})^\top \bm{\Sigma}^{-1}(\bm{y} - K_c\bm{\beta}), \\
    s_\beta(\bm{y}, \bm{\gamma}) &= (K_c\bm{\gamma})^\top \bm{\Sigma}^{-1}(\bm{y} - K_c\bm{\beta}).
\end{align}
The efficient score is $s_{c,\text{eff}}(\bm{y}) = s_c(\bm{y}) - s_\beta(\bm{y}, \bm{\gamma}^*)$, 
where, in the unconstrained case,
\begin{equation}
\bm{\gamma}^*=
\operatorname*{argmin}_{\bm{\gamma}\in\mathbb{R}^m}\,
(\dot K_c\bm{\beta}-K_ch)^\top
\bm{\Sigma}^{-1}
(\dot K_c\bm{\beta}-K_ch).
\end{equation}
Provided that $K_c^\top\bm{\Sigma}^{-1}K_c$ is invertible, the solution is
\begin{equation}
\bm{\gamma}^*
=
(K_c^\top\bm{\Sigma}^{-1}K_c)^{-1}
K_c^\top\bm{\Sigma}^{-1}\dot K_c\bm{\beta}.
\end{equation}
In the constrained case, admissible nuisance directions must preserve the normalization of $f$, and therefore satisfy
\begin{equation}
w^\top \bm{\gamma}=0.
\end{equation}
The constrained projection problem is thus
\begin{equation}
\bm{\gamma}^*
=
\operatorname*{argmin}_{\bm{\gamma}:\,w^\top \bm{\gamma}=0}
(\dot K_c\bm{\beta}-K_c\bm{\gamma})^\top
\bm{\Sigma}^{-1}
(\dot K_c\bm{\beta}-K_c\bm{\gamma}).
\end{equation}
Setting $U := 2K^\top_c \bm{\Sigma}^{-1} K_c$ and $b := 2K^\top_c \bm{\Sigma}^{-1} \dot{K}_c\bm{\beta}$, the optimality conditions form the system
\begin{equation}
\label{eq:KKT}
    \begin{bmatrix} U & w \\ w^\top & 0 \end{bmatrix}
    \begin{bmatrix} \bm{\gamma}^* \\ \lambda \end{bmatrix}
    =
    \begin{bmatrix} b \\ 0 \end{bmatrix},
\end{equation}
which can be solved for $\bm{\gamma}^*$. The efficient score and EIF then take the same form as \eqref{eq:efficient_score} and \eqref{eq:EIF} with $f$ replaced by $\bm{\beta}$:
\begin{equation}
    s_{c,\text{eff}}(\bm{y}) = (\dot{K}_c\bm{\beta} - K_c\bm{\gamma}^*)^\top {\bm\Sigma}^{-1}(\bm{y} - K_c\bm{\beta}),
\end{equation}
\begin{equation}
    \varphi(\bm{y};c,\bm{\beta}) = \frac{(\dot{K}_c\bm{\beta} - K_c\bm{\gamma}^*)^\top {\bm\Sigma}^{-1}(\bm{y} - K_c\bm{\beta})}
    {(\dot{K}_c\bm{\beta} - K_c\bm{\gamma}^*)^\top {\bm\Sigma}^{-1}(\dot{K}_c\bm{\beta} - K_c\bm{\gamma}^*)}.
\end{equation}
On the computational side, the implementation is straightforward. Notice that the design matrix $A$ only needs to be computed once, and it can be directly computed using the FK table with the basis functions evaluated at the specified grid points. Specifically, we replace the PDF $f$ in Eq.\,\eqref{eq:FKsum} with $B_k(y)/y$ for each basis function $B_k$, which gives the corresponding columns of $A$. Once $A$ is constructed, the matrices $K_c$ and $\dot{K}_c$ can be obtained efficiently by element-wise multiplying the rows of $A$ with $\bm{\nu}(c)$ and $\bm{\nu}'(c)$, respectively. Thus, most quantities required for the computation can be formed with simple matrix operations. The main remaining computational task is to solve the least squares problem to obtain $h^*$, which can be done efficiently using standard linear algebra routines. 
\section{Application: $T_3$ from BCDMS Data}
\label{sec:application_BCDMS}
As a first example, we apply the semiparametric estimation methodology described in Sect.~\ref{sec:methodology} to the case of extracting Wilson coefficients in a 
controlled setup based on simulated DIS data from the BCDMS experiment~\cite{BCDMS:1989ggw, BCDMS:1989qop} sensitive to the non-singlet triplet PDF $T_3$. 
We will show that our method can accurately estimate EFT parameters even when starting from a preliminary estimate based on SM PDFs. For the estimates in this section and Sect.~\ref{sec:application_HERA}, we use $m=8$ B-spline functions, as we found that this provides a reasonable compromise between flexibility and stability.

Both applications presented in this section and in Sect.~\ref{sec:application_HERA} are deliberately restricted to a single PDF combination and a single experimental dataset.
While a realistic global PDF analysis involves simultaneously constraining 7 or 8 flavour combinations across hundreds of correlated measurements from a variety of processes, 
this controlled setting serves to validate the semiparametric method before extending to the global case. 
Crucially, the problem remains genuinely non-trivial even in this simplified setting: the PDF is still an infinite-dimensional nuisance parameter, so standard parametric 
inference does not apply; the EFT correction enters multiplicatively through the partonic cross section (see Sect.~\ref{sec:model}); and the one-step estimator must invert 
a realistic experimental covariance matrix with hundreds of correlated data points.
The choice of $T_3$ as the single PDF combination is itself physically motivated --- as we show below, the structure function difference $F_2^p - F_2^d$ isolates $T_3$ exactly 
from the remaining flavours, making it a clean, natural single-PDF observable rather than an artificial construction.
Single-flavour determinations of this kind have historically served as a testbed for new methodologies before their extension to global analyses: the NNPDF approach was first 
applied to the non-singlet distribution from DIS data~\cite{DelDebbio:2007ee}, then to a complete DIS fit~\cite{Ball:2008by}, before evolving into the current global analyses that 
include data from proton-(anti)proton colliders~\cite{NNPDF:2021njg}.
Following the same strategy, $T_3$ has been used as a case study for Gaussian-process-based PDF determinations~\cite{Candido:2024hjt}, for studying the neural tangent kernel 
(NTK) of neural-network PDF fits~\cite{Chiefa:2025cap}, and for estimating the faithfulness of machine-learning approaches to inverse problems~\cite{Elmer:2026sbp}.
These applications are therefore preparatory to a realistic global analysis, as discussed in Sect.~\ref{sec:conclusion}.

\begin{figure}[htbp]
    \centering
    \includegraphics[width=0.9\textwidth]{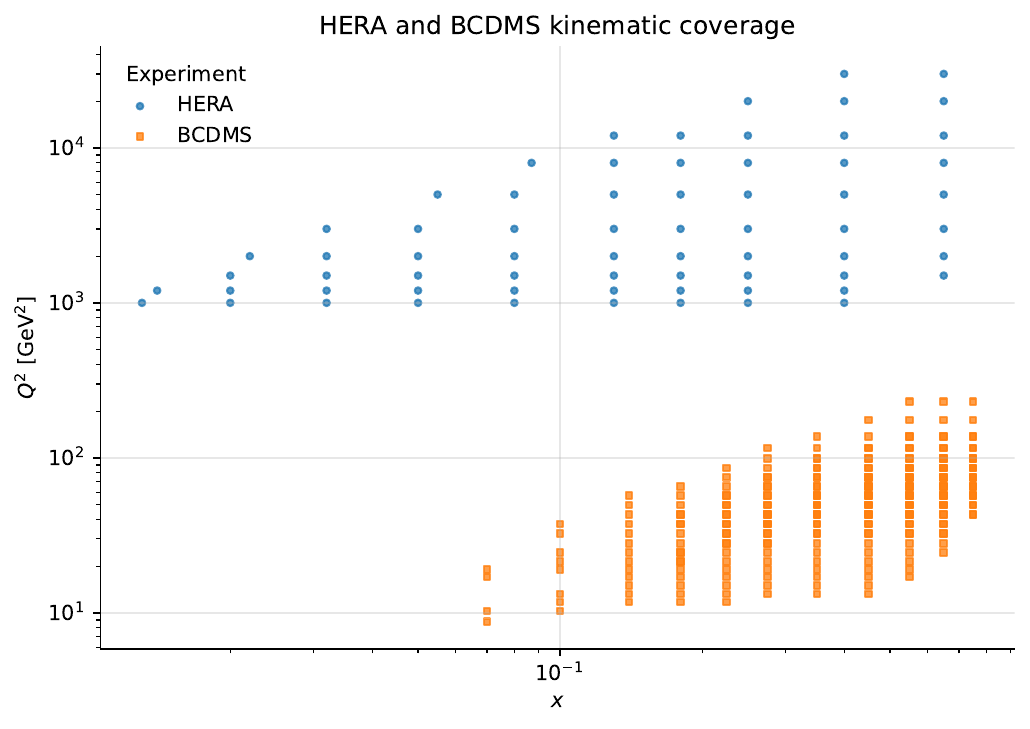}
    \caption{Kinematic coverage of the BCDMS~\cite{BCDMS:1989ggw, BCDMS:1989qop} and HERA\cite{H1:2015ubc} data used in the analysis. 
    Each point corresponds to a data point with coordinates given by $(x, Q^2)$.}
    \label{fig:kinematic-coverage}
\end{figure}

\subsection{Pseudo-Data Generation}
\label{subsec:pseudodata_bcdms}
We consider the fixed-target neutral current (NC) structure functions $F_2^p$ and $F_2^n$ as measured by the BCDMS experiment. The proton structure function $F_2^p$ 
can be described at NNLO in QCD by a sum of convolutions of the $g, \Sigma, T_3, T_8$ and $T_{15}$ PDF flavours in the evolution basis \cite{Candido:2024hjt},
\begin{equation}
    F_2^p = C_g \otimes g + C_{\Sigma} \otimes \Sigma + C_{T_3} \otimes T_3 + C_{T_8} \otimes T_8 + C_{T_{15}} \otimes T_{15} \, .
    \label{eq:F2p}
\end{equation}
Similarly, assuming isospin symmetry between the up and down quarks, the neutron structure function $F_2^n$ can be written as 
\begin{equation}
    F_2^n = C_g \otimes g + C_{\Sigma} \otimes \Sigma - C_{T_3} \otimes T_3 + C_{T_8} \otimes T_8 + C_{T_{15}} \otimes T_{15} \, ,
\end{equation}
which differs from the expression for $F_2^p$ in Eq.~\eqref{eq:F2p} by an opposite sign in front of the coefficient function $C_{T_3}$ due to the asymmetry between the $u^+$ and $d^+$ distributions entering $T_3$. For a deuterium target, being a bound state of a proton and neutron, we therefore find that the contribution coming from $T_3$ to its structure function cancels, leaving
\begin{equation}
    F_2^d = C_g \otimes g + C_\Sigma \otimes \Sigma + C_{T_8} \otimes T_8 + C_{T_{15}} \otimes T_{15} \, .
    \label{eq:F2d}
 \end{equation}
The difference between Eq.~\eqref{eq:F2d} and \eqref{eq:F2p} thus isolates $T_3$,
\begin{equation}
    F_2^p - F_2^d = C_{T_3} \otimes T_3 \, ,
\end{equation}
making it an ideal proving ground to test our approach in a simple setting where only a single PDF combination enters. Indeed, the corresponding SM predictions ${\bf T}_{\rm SM}(f)$ can be obtained by multiplying $T_3$ with the FK tables in the evolution basis,
\begin{equation}
    T_{\rm SM}(x_i, Q_i^2; f) = \sum_{\alpha=1}^{n_{\rm grid}}
    ({\rm FK})_{T_3\,\alpha}(x_i, x_\alpha, Q_i^2, Q_0^2)\,
    T_3(x_\alpha, Q_0^2) \, ,
\end{equation}
and the EFT predictions follow straightforwardly in our model as
\begin{equation}
    {\bf T}(c, f) = {\bm \nu}(c) \odot {\bf T}_{\rm SM}(f) \, .
\end{equation}

We provide in Table \ref{tab:bcmds_dataset} an overview of the BCDMS measurements of $F_2^p$ and $F_2^n$ we include in our fit, indicating for each its publication reference, number of data points, Bj\"orken $x$ range, $Q^2$ range, and finally the source of the corresponding theoretical prediction. Both structure functions $F_2^p$ and $F_2^d$ are measured at Bj\"orken $x$ values between $x=0.07$ and $x=0.75$, with corresponding $Q$ values ranging between $Q=2.7$ GeV and $Q=15.1$ GeV, and $Q=3.0$ GeV and $Q=15.1$ GeV for the proton and deuteron, respectively. Focusing on their overlap in the $(x, Q^2)$ plane to construct the observable $F_2^p - F_2^d$, and applying the NNPDF4.0 data set selection criteria \cite{ball2022}, we find that 254 data points remain. We show the corresponding kinematic coverage in Fig.~\ref{fig:kinematic-coverage}. Theoretical predictions are computed with {\sc YADISM} \cite{Candido:2024rkr} at NNLO in QCD. 

\begin{table}[htbp]
\centering
\caption{BCDMS fixed-target DIS datasets included in the analysis.}\smallskip
\begin{tabular}{cccccc}
\hline
Obs. & Ref. & $N_{\rm dat}$ & $x$ range & $Q$ range (GeV) & Theory \\
\hline
$F_2^p$ & \cite{BCDMS:1989ggw} & 351  & $0.07 \leq x \leq 0.75$ & $2.7 \leq Q \leq 15.1$ & YADISM \cite{Candido:2024rkr} \\
$F_2^d$ & \cite{BCDMS:1989qop} & 254  & $0.07 \leq x \leq 0.75$ & $3.0 \leq Q \leq 15.1$ & YADISM \cite{Candido:2024rkr} \\
\hline
\end{tabular}
\label{tab:bcmds_dataset}
\end{table}

Next, we generate pseudo-data $\bm{y}$ from a known underlying law by adding experimental
Gaussian noise to the theoretical prediction ${\bf T}(c,f)$:
\begin{equation}
     \bm{y} \sim \mathcal{N}\!\bigl(\mathbf{T}(c,f),\, \bm{\Sigma}\bigr)\, ,
     \label{eq:pseudodata_bcdms}
\end{equation}
where the covariance matrix $\boldsymbol{\Sigma}$ is the experimental covariance
of the difference $F_2^p - F_2^d$,
\begin{equation}
    \bm{\Sigma} = {\rm Cov}[F_2^p-F_2^d] = {\rm Cov}[F_2^p] + {\rm Cov}[F_2^d] - {\rm Cov}[F_2^p, F_2^d]  -  {\rm Cov}[F_2^d, F_2^p] \, .
\end{equation}
In the remainder of this section, we adopt pseudo-data generated from Eq.~\eqref{eq:pseudodata_bcdms} at $c=500$ and $\Lambda = 500~\mathrm{GeV}$, including both linear and quadratic EFT corrections. This seemingly high value of $c$ is chosen in order to induce a sufficiently pronounced BSM effect given our default value of $\Lambda$. However, as argued in Sect.~\ref{subsec:eft_toy_example}, $c$ and $\Lambda$ enter only through the combination $c/\Lambda^2$, so this choice is equivalent, for example, to setting $\Lambda = 100~\mathrm{GeV}$ and $c=20$. 

\begin{figure}[htbp]
    \centering
    \includegraphics[width=\textwidth]{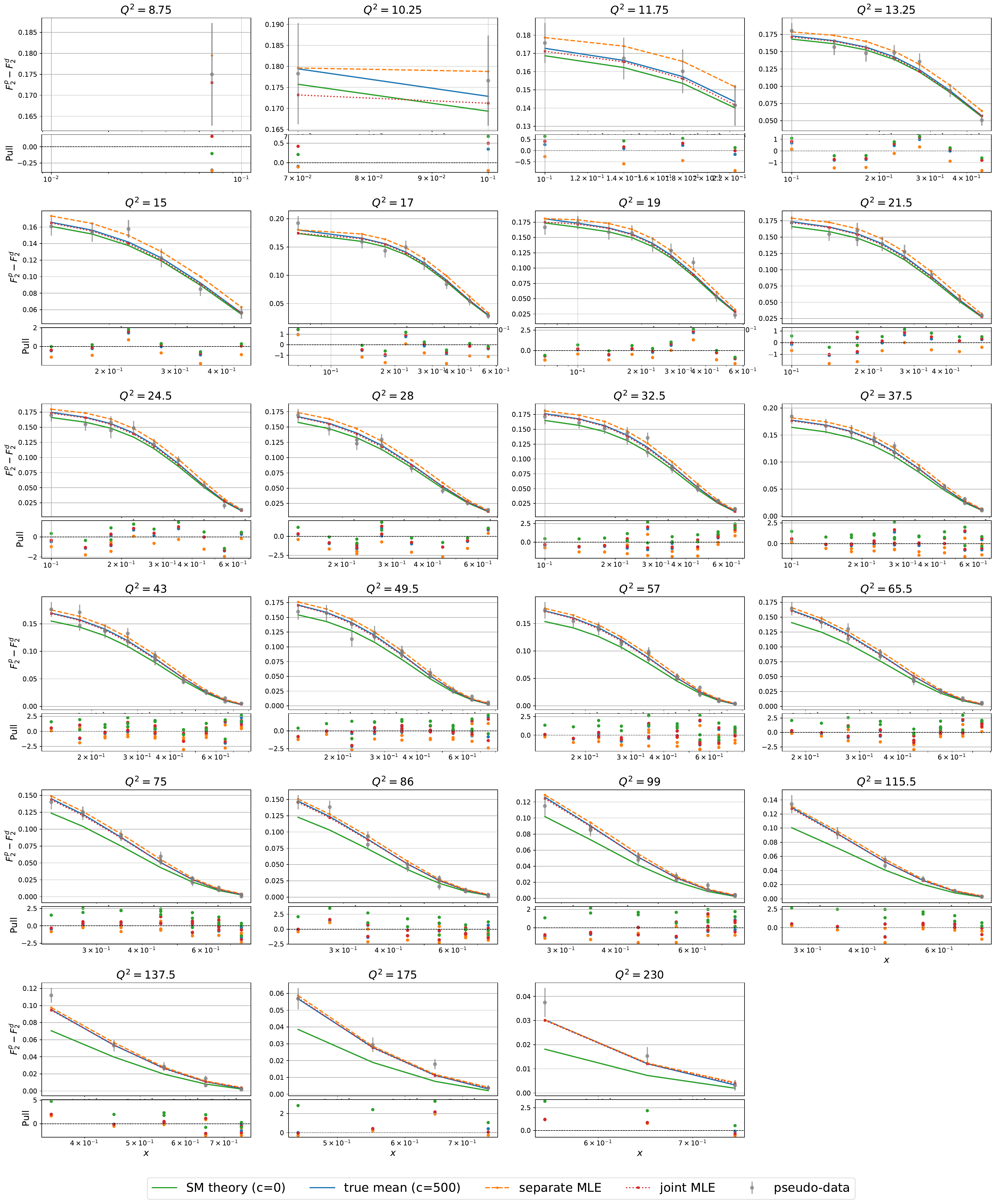}
    \caption{Data-level comparison between the $F_2^p-F_2^d$ structure function pseudo-data and predictions obtained from separate and joint fits of the Wilson coefficient $c$ and the PDF $f$. Each panel corresponds to a different $Q^2$ bin. The gray points with error bars represent the pseudo-data. The solid green curve shows the SM prediction ($c=0$), while the solid blue curve shows the true mean corresponding to $c=500$. The orange dashed curve is the fitted prediction obtained from the separate maximum likelihood estimates $\hat{c}_0$ and $\hat{f}_0$, whereas the red dotted curve is the fitted prediction obtained from the joint maximum likelihood estimates of $c$ and $f$. The lower subpanels show the corresponding pulls relative to the pseudo-data.} 
    \label{fig:BCDMS_data-fit}
\end{figure}

\medskip

\subsection{Results}
We now present the main results of applying semiparametric inference in the context of BCDMS data. The fitting procedure is as described in Sect.~\ref{sec:methodology}, while the pseudo-data is generated as described in Sect.~\ref{subsec:pseudodata_bcdms}.

Fig.~\ref{fig:BCDMS_data-fit} presents the results of the fit to $F_2^p-F_2^d$ at the data level, with each panel corresponding to a different $Q^2$ bin. The preliminary fitted curve (orange) closely follows the true BSM curve (blue), particularly in the high-$Q^2$ bins, where the BSM contribution is more pronounced. This behavior is not surprising since the preliminary estimates are obtained by fitting data that already contain the BSM contribution. Consequently, the preliminary fit can absorb part of the BSM-induced structure through the fitted PDF and reproduce the overall shape of the observed data. However, good agreement at the data level does not necessarily imply an accurate estimate of the parameter $c$. In particular, when the PDF is estimated under the SM-based preliminary fitting procedure, part of the BSM contribution can be absorbed into the estimated PDF, leading to bias in the resulting estimate of $c$.

\begin{figure}[htbp]
    \centering
    \includegraphics[width=\textwidth]{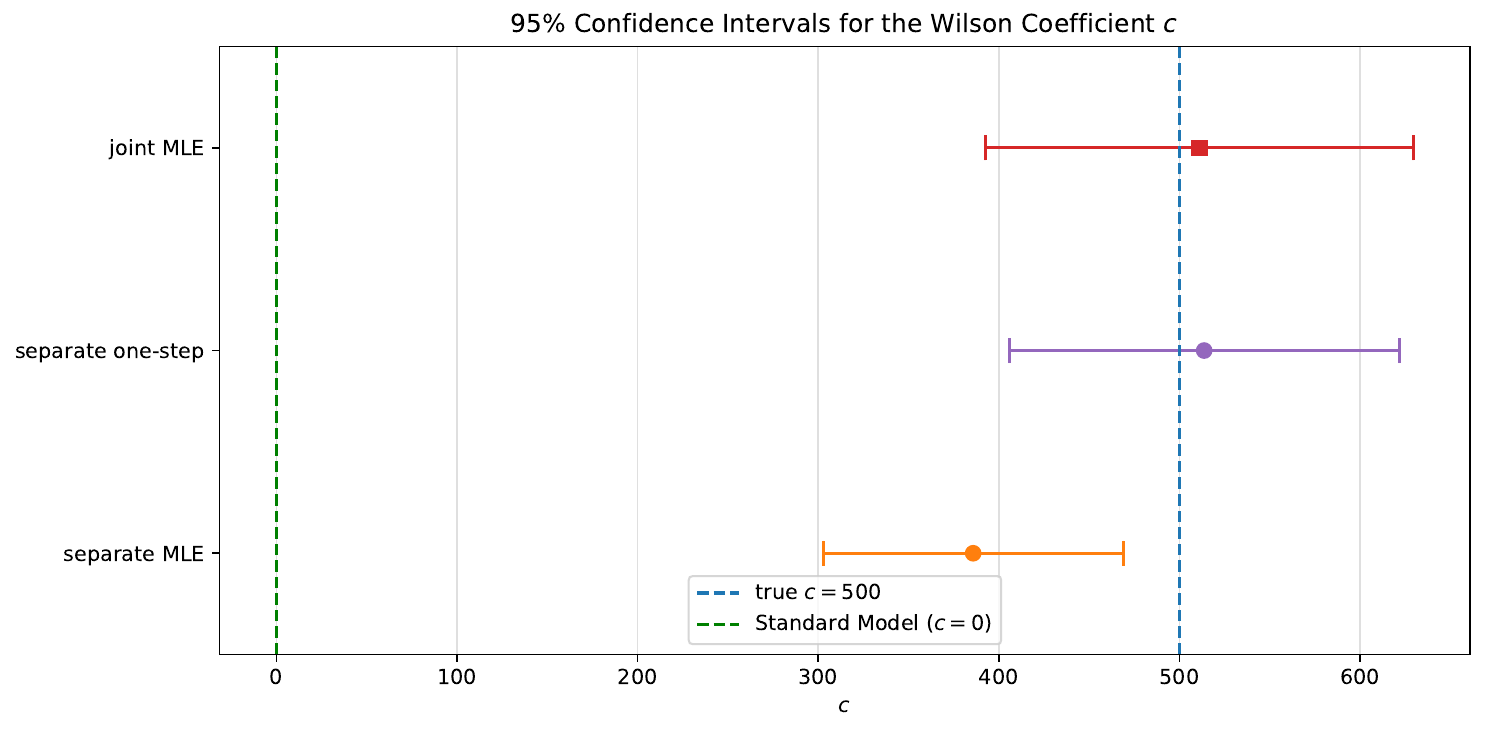}
    \caption{95\% confidence intervals for the Wilson coefficient $c$. The blue dashed line indicates the true value $c=500$, while the green dashed line indicates the Standard Model value $c=0$. The orange interval corresponds to the preliminary estimate $\hat{c}_0$ obtained by separate maximum likelihood estimation, the purple interval corresponds to its one-step estimate $\hat{c}_{\rm one\text{-}step}$, and the red interval corresponds to the joint maximum likelihood estimate of $c$. The markers within the intervals indicate the corresponding point estimates.}
    \label{fig:BCDMS_CI_c}
\end{figure}

\begin{figure}[htbp]
    \centering
    \includegraphics[width=0.85\textwidth]{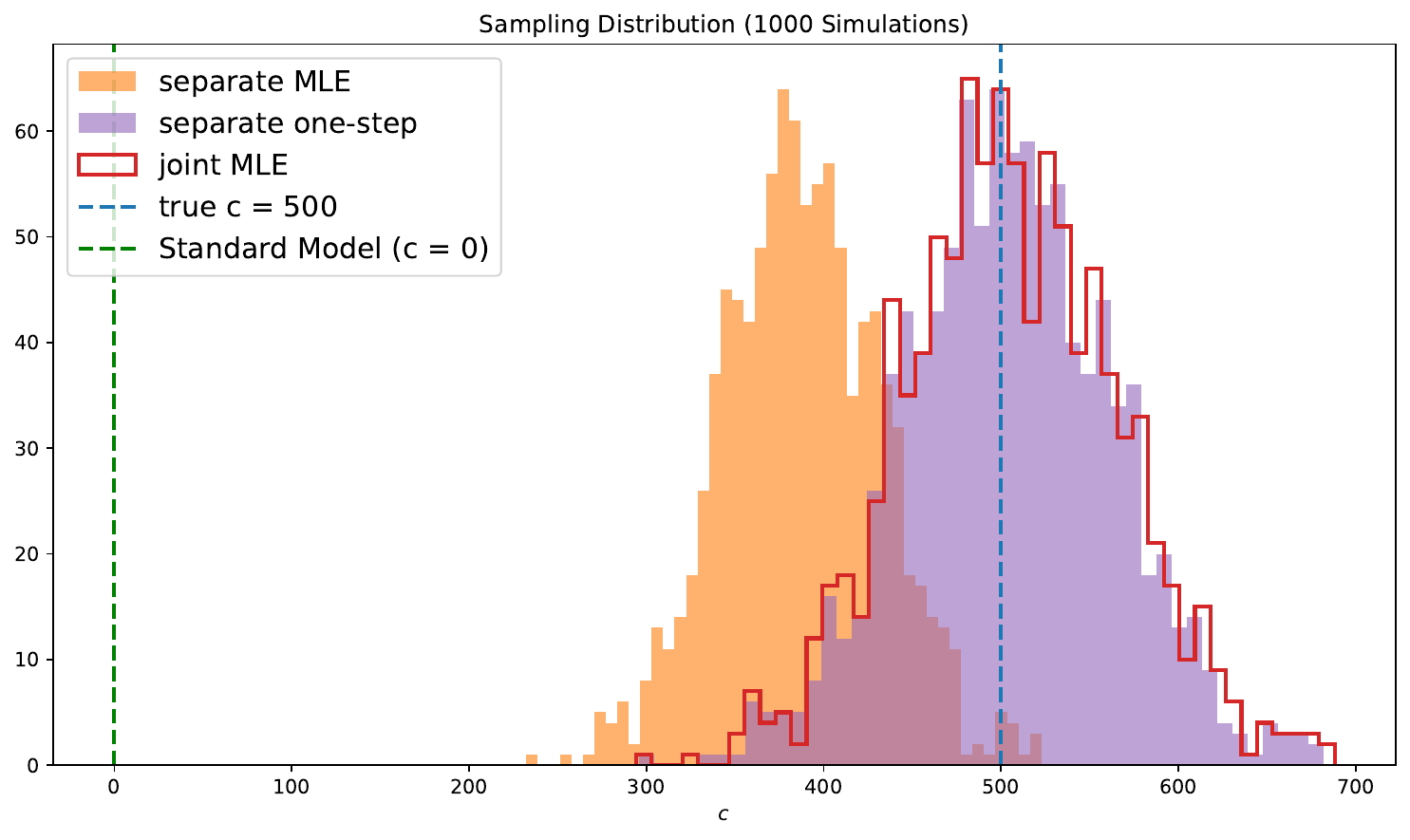}
    \caption{Sampling distributions of point estimates of the Wilson coefficient $c$. The data-generation and estimation procedure is repeated 1000 times. The filled orange histogram shows the sampling distribution of the preliminary separate MLE $\hat{c}_0$, the purple histogram shows the sampling distribution of the one-step estimator $\hat{c}_{\rm one\text{-}step}$, and the red outlined histogram shows the sampling distribution of the joint MLE of $c$. The vertical blue dashed line indicates the true value $c=500$, while the green dashed line indicates the Standard Model value $c=0$.}
    \label{fig:BCDMS_sampling_dist_c}
\end{figure}

Indeed, the preliminary estimate $\hat{c}_0$ is biased and systematically underestimates the true value of $c=500$. The lower orange line in Fig.~\ref{fig:BCDMS_CI_c} shows $\hat{c}_0$ and its corresponding 95\% confidence interval for $c$. The confidence interval is computed using the parametric bootstrap method with 500 bootstrap samples. 
We estimate the standard error of $\hat{c}_0$ by the standard deviation of the bootstrap estimates of $c$. The bootstrap confidence interval is then given by $\hat{c}_0 \pm z_{\alpha/2} \hat{\sigma}$, where $\hat{\sigma}$ is the estimated standard error. As shown in Fig.~\ref{fig:BCDMS_CI_c}, the resulting interval is shifted below the true value and fails to cover $c=500$. In contrast, the one-step estimator $\hat{c}_{\rm one-step}$ is much closer to the true value. Its confidence interval is constructed using the asymptotic normality of the one-step estimator; see Eq.~\eqref{eq:ci_one_step}. The resulting 95\% confidence interval successfully covers the true value $c=500$. For comparison, we also compute the joint MLE by simultaneously optimizing over $c$ and the PDF rather than first estimating the PDF through the preliminary fitting procedure. In this case, the joint MLE point estimate is close to the one-step estimate, and its corresponding 95\% confidence interval, constructed using the asymptotic normality of the MLE with the standard error estimated from the inverse Hessian of the negative log-likelihood, also covers the true value.

Furthermore, we perform a simulation study in which the full data-generation and estimation procedure is repeated 1000 times. For each simulated dataset, we compute the preliminary estimate $\hat{c}_0$, the one-step estimate $\hat{c}_{\rm one-step}$, and the joint MLE. Fig.~\ref{fig:BCDMS_sampling_dist_c} shows the resulting sampling distributions. The sampling distribution of $\hat{c}_0$ is centered below the true value of $c$, confirming that the preliminary estimator is systematically biased downward. In contrast, the sampling distribution of $\hat{c}_{\rm one-step}$ is centered much closer to the true value, which shows that the one-step correction effectively removes most of this bias. The sampling distribution of the joint MLE is likewise centered near the true value and closely resembles that of the one-step estimator. 

\begin{figure}[htbp]
    \centering
    \includegraphics[width=\textwidth]{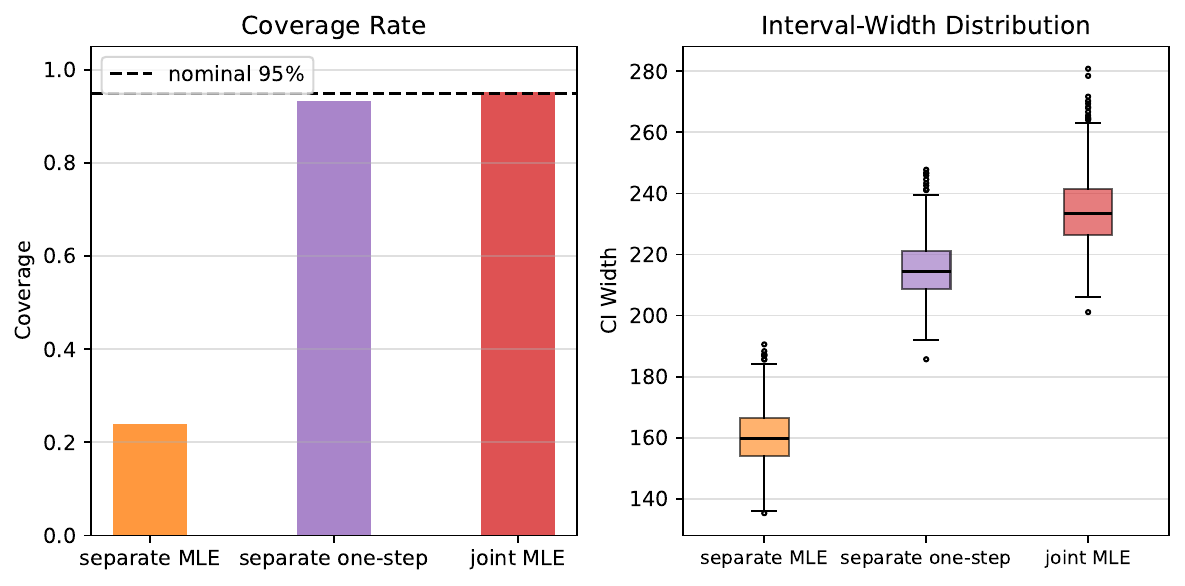}
    \caption{Coverage rates and confidence interval widths for the Wilson coefficient $c$. The coverage rate is estimated as the proportion of the 1000 simulated confidence intervals that contain the true value $c=500$, with the black dashed line indicating the nominal 95\% coverage level. The right panel shows the distributions of confidence-interval widths, computed as the difference between the upper and lower interval bounds. Results are shown for the separate MLE $\hat{c}_0$, the one-step estimator $\hat{c}_{\rm one\text{-}step}$, and the joint MLE of $c$.}
    \label{fig:BCDMS_CI_coverage_width}
\end{figure}

Fig.~\ref{fig:BCDMS_CI_coverage_width} summarizes the empirical coverage and widths of the confidence intervals for $c$ over the 1000 simulations. The coverage rate is estimated as the proportion of simulated confidence intervals containing the true value $c=500$, while the interval width is defined as the difference between its upper and lower endpoints. The confidence intervals based on the preliminary estimator $\hat{c}_0$ have coverage substantially below the nominal 95\% level, consistent with the systematic bias observed in its sampling distribution. In comparison, the intervals based on the one-step estimator achieve coverage much closer to the nominal level. The joint-MLE intervals similarly provide substantially improved coverage relative to those based on the preliminary estimator. In terms of width, we can see that the confidence intervals based on $\hat{c}_{\rm one-step}$ are wider than those based on $\hat{c}_0$. This increase should not be interpreted as a deficiency of the one-step procedure. The preliminary intervals are narrower in part because they do not adequately account for the uncertainty associated with estimating the nuisance parameter, which contributes to their substantial undercoverage. The one-step correction incorporates the effect of nuisance estimation through the influence-function correction, resulting in uncertainty quantification that more accurately reflects the uncertainty of the estimator. In this example, the one-step intervals are narrower than the corresponding joint-MLE intervals while achieving coverage close to the nominal level. Moreover, under regularity conditions, the one-step estimator is asymptotically efficient, indicating that it has the smallest asymptotic variance among regular estimators. 

\begin{figure}[htbp]
    \centering
    \includegraphics[width=0.8\textwidth]{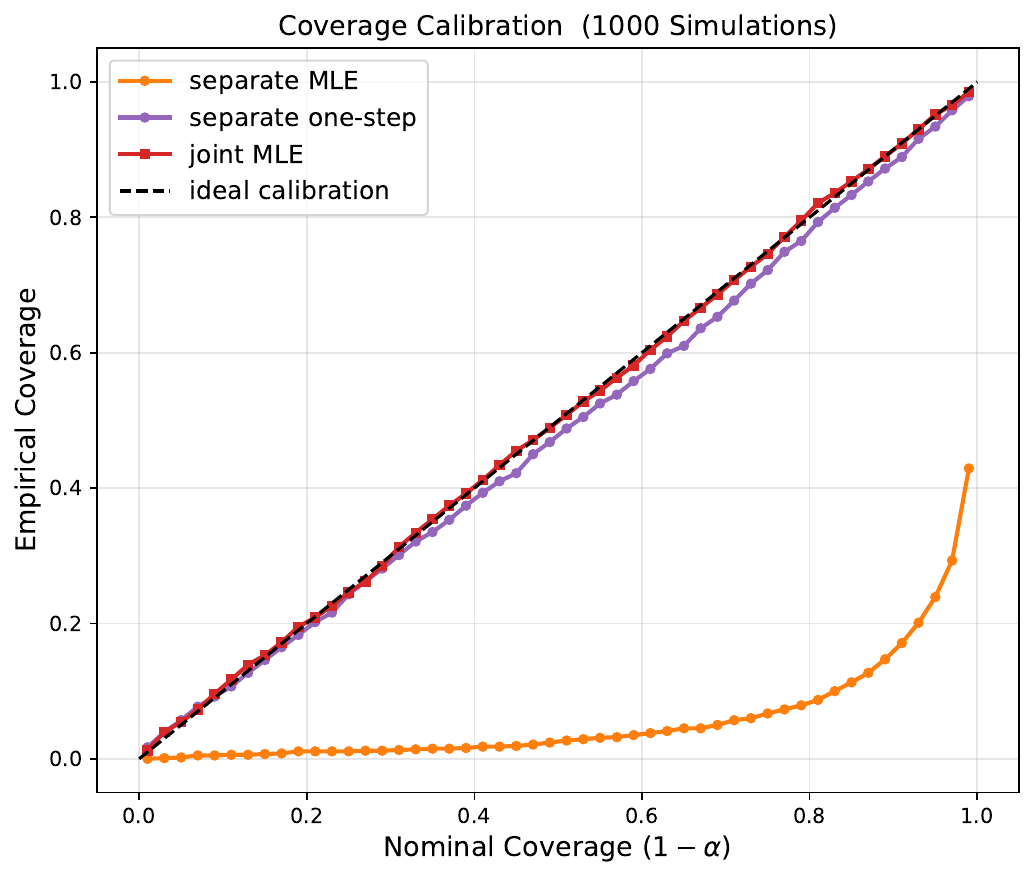}
    \caption{Coverage calibration of the confidence intervals for the Wilson coefficient $c$. The horizontal axis represents the nominal confidence level, while the vertical axis shows the empirical coverage rate estimated from 1000 independent simulations. The orange curve with circular markers corresponds to confidence intervals based on the separate MLE $\hat{c}_0$, the purple curve corresponds to intervals based on the one-step estimator $\hat{c}_{\rm one\text{-}step}$, and the red curve with square markers corresponds to intervals based on the joint MLE of $c$. The black dashed diagonal line represents ideal calibration, for which the empirical coverage matches the nominal confidence level.}
    \label{fig:BCDMS_coverage_calibration}
\end{figure}

Finally, we assess the coverage calibration of the confidence intervals for $c$ by plotting the empirical coverage rate against the nominal confidence level. Fig.~\ref{fig:BCDMS_coverage_calibration} shows that the confidence intervals based on $\hat{c}_0$ are poorly calibrated, with empirical coverage rates substantially below the nominal levels. By contrast, the confidence intervals based on the one-step estimator track the diagonal much more closely, indicating substantially improved calibration that is comparable to the calibration of the joint MLE.

\section{Application: $V$ from HERA Data}
\label{sec:application_HERA}
As a second application, we consider the extraction of the Wilson coefficient $c$ from the NC structure function $F_3^{\gamma Z}(x, Q^2)$ as measured by HERA~\cite{H1:2015ubc}. Unlike the BCDMS experiment in Sect.~\ref{sec:application_BCDMS}, the proton structure function $F_3^{\gamma Z}$ is predominantly sensitive to the valence quark distributions. As these are subject to the valence sum rules, the goal of this section is to illustrate how our methodology can be extended to hold also in the presence of normalization constraints. 

\subsection{Pseudo-Data Generation}
In NC DIS, the $F_3$ structure function receives contributions from photon-$Z$ interference and $Z$ exchange and can be written in terms of electroweak parameters as \cite{H1:2015ubc}
\begin{equation}
    F_3(x, Q^2) = - \kappa_Z a_e F_3^{\gamma Z}(x, Q^2) + 2\kappa_Z^2 v_ea_e F_3^Z(x, Q^2) \, ,
\end{equation}
where $v_e$ and $a_e$ denote respectively the vector and axial-vector couplings of the electron to the $Z$-boson with mass $M_Z$, while the kinematic prefactor $\kappa_Z$ is defined in terms of the weak mixing angle $\theta_W$ as
\begin{equation}
    \kappa_Z(Q^2) = \frac{Q^2}{(Q^2+M_Z^2)(4\sin^2\theta_W \cos^2\theta_W)}\, .
\end{equation}
In the quark-parton model, which assumes gluons are not present and the longitudinal contribution $F_L=0$, the contribution $F_3^{\gamma Z}$ coming from photon-$Z$ interference can be approximated in terms of the valence up and down distributions $u_V$ and $d_V$,
\begin{equation}
    F_3^{\gamma Z}(x, Q^2) \approx 2 \left(e_ua_u u_V + e_da_d d_V\right),
    \label{eq:F3_qpm}
\end{equation}
where $e_u=2/3$ and $e_d=-1/3$ denote the electric charge of the up and down-type quarks, while $a_u=1/2$ and $a_d=-1/2$ denote the axial vector coupling of the up and down-type quark to the $Z$-boson, respectively. Eq.~\eqref{eq:F3_qpm} thus takes the form
\begin{equation}
    F_3^{\gamma Z}(x, Q^2) \approx \frac{1}{3}(2u_V + d_V ) \, .
    \label{eq:F3_qpm_flavour_basis}
\end{equation}
It then follows straightforwardly that Eq.~\eqref{eq:F3_qpm_flavour_basis} can be written in terms of the evolution basis as 
\begin{equation}
    F_3^{\gamma Z}(x, Q^2) \approx \frac{4}{9}V + \frac{1}{6}V_3 + \frac{1}{18}V_8 \, ,
    \label{eq:F3_qpm_evol_basis}
\end{equation}
which indicates how the dominant sensitivity enters through the total valence distribution~$V$. Beyond leading order, the naive quark-parton model no longer holds, and Eq.~\eqref{eq:F3_qpm_evol_basis} receives additional contributions coming from quark-gluon radiation. However, these contributions are small in comparison and its dominant sensitivity to $V$ remains. The $F_3^{\gamma Z}$ structure function therefore provides a reasonable probe to constrain $V$ also beyond leading order. 

\begin{figure}[htbp]
    \centering
    \includegraphics[width=\textwidth]{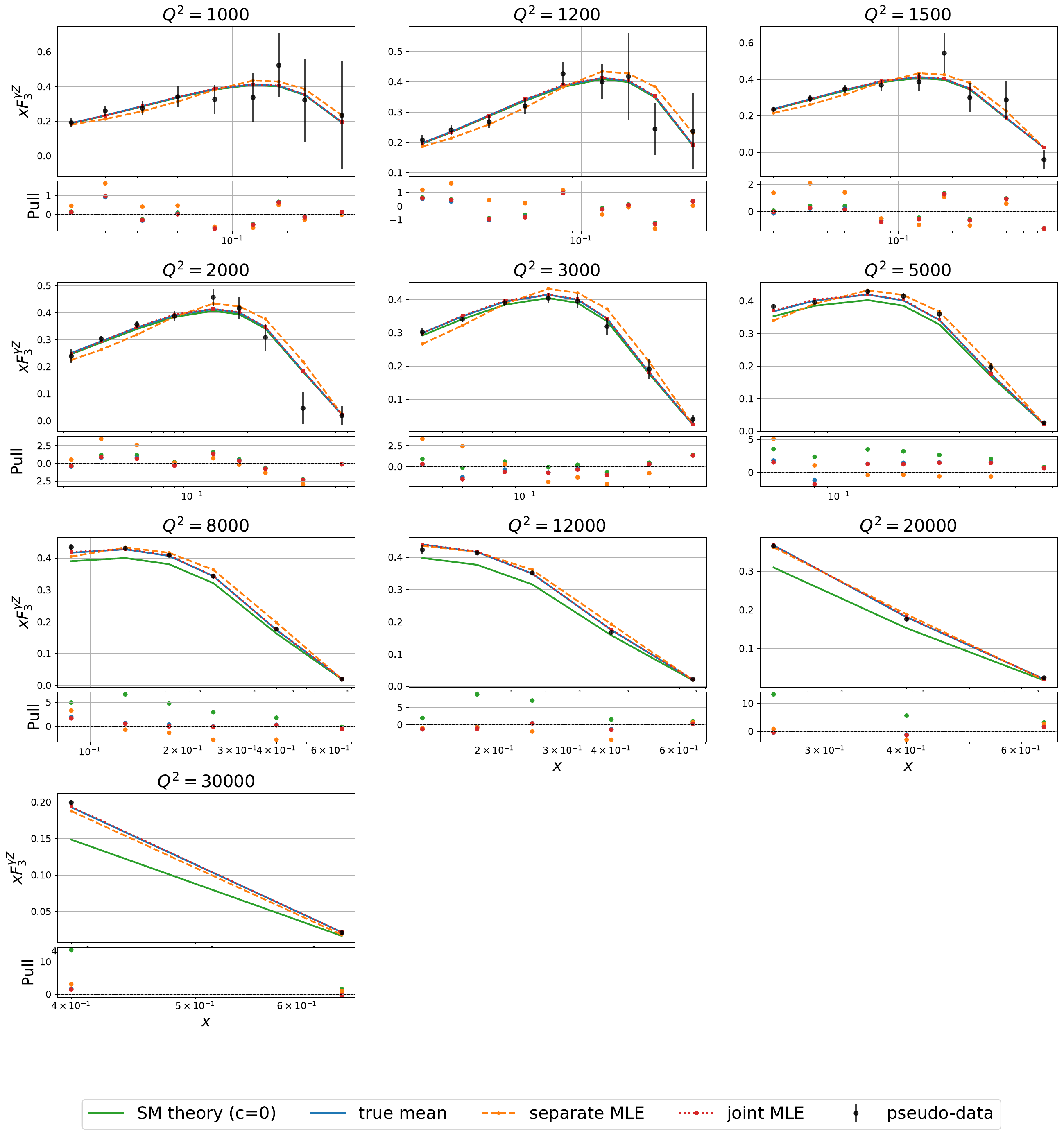}
    \caption{Data-level comparison between the $F_3^{\gamma Z}$ structure function pseudo-data and predictions obtained from separate and joint fits of the Wilson coefficient $c$ and the PDF $f$. Each panel corresponds to a different $Q^2$ bin. The gray points with error bars represent the pseudo-data. The solid green curve shows the Standard Model prediction ($c=0$), while the solid blue curve shows the true mean corresponding to $c=2$. The orange dashed curve is the fitted prediction obtained from the separate maximum likelihood estimates $\hat{c}_0$ and $\hat{f}_0$, whereas the red dotted curve is the fitted prediction obtained from the joint maximum likelihood estimates of $c$ and $f$. The lower subpanels show the corresponding pulls relative to the pseudo-data.}
    \label{fig:HERA_data-fit}
\end{figure}

Regarding pseudo-data generation, we consider the HERA measurement of the $F_3^{\gamma Z}$ structure function from~\cite{H1:2015ubc} consisting of 67 data points in $(x, Q^2)$ separated in 10 distinct $Q^2$ bins ranging from $Q^2=1000$ GeV$^2$ to $Q^2=30000$ GeV$^2$, with $x$ ranging from $x=0.013$ to $x=0.650$. We display this kinematic coverage in Fig.~\ref{fig:kinematic-coverage}.  Following the procedure of Sect.~\ref{subsec:eft_toy_example}, we generate pseudo-data for this dataset with the SM prediction ${\bf T}_{\rm SM}$ computed at NNLO
in QCD with \textsc{YADISM}~\cite{Candido:2024rkr} and the EFT correction ${\bm \nu}(c)$ evaluated at $c=2$. The corresponding uncertainties follow the experimental covariance matrix of the HERA $F_3^{\gamma Z}$ measurement. As $F_3^{\gamma Z}$ is dominantly sensitive to $V$, we fit only $V$ and fix all other contributions to a pre-determined PDF set, which we take as NNPDF4.0 \cite{ball2022}. In addition, we fix $V$ to NNPDF4.0 outside the data region so that we only fit in a region of $x$ where the data provides sufficient sensitivity. We have verified explicitly that without this additional constraint the forward map becomes ill-conditioned and unstable.

\begin{figure}[htbp]
    \centering
    \includegraphics[width=\textwidth]{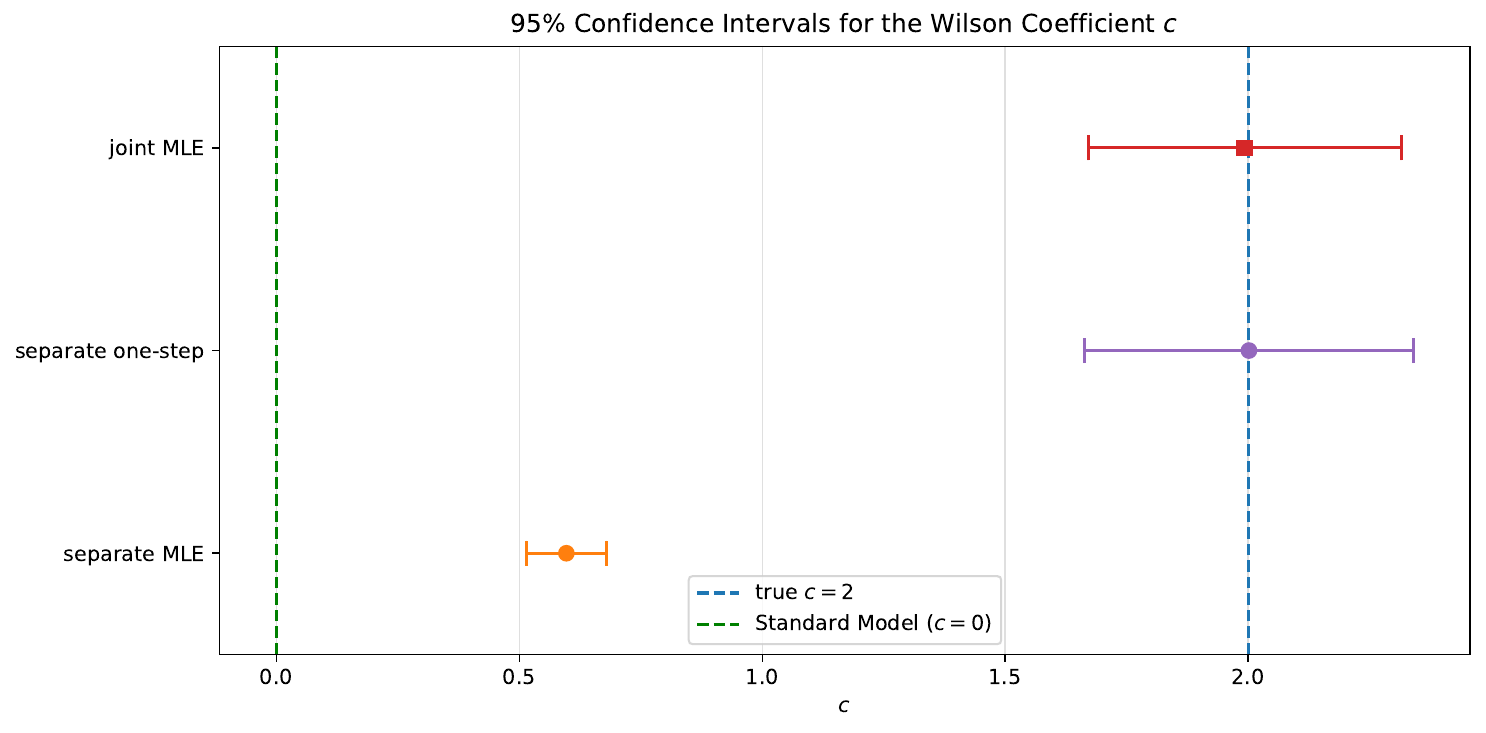}
    \caption{95\% confidence intervals for the Wilson coefficient $c$. The blue dashed line indicates the true value $c=2$, while the green dashed line indicates the Standard Model value $c=0$. The orange interval corresponds to the preliminary estimate $\hat{c}_0$ obtained by separate maximum likelihood estimation, the purple interval corresponds to its one-step estimator $\hat{c}_{\rm one\text{-}step}$, and the red interval corresponds to the joint maximum likelihood estimator of $c$. The markers within the intervals indicate the corresponding point estimates.}
    \label{fig:HERA_CI_c}
\end{figure}

\begin{figure}[htbp]
    \centering
    \includegraphics[width=0.85\textwidth]{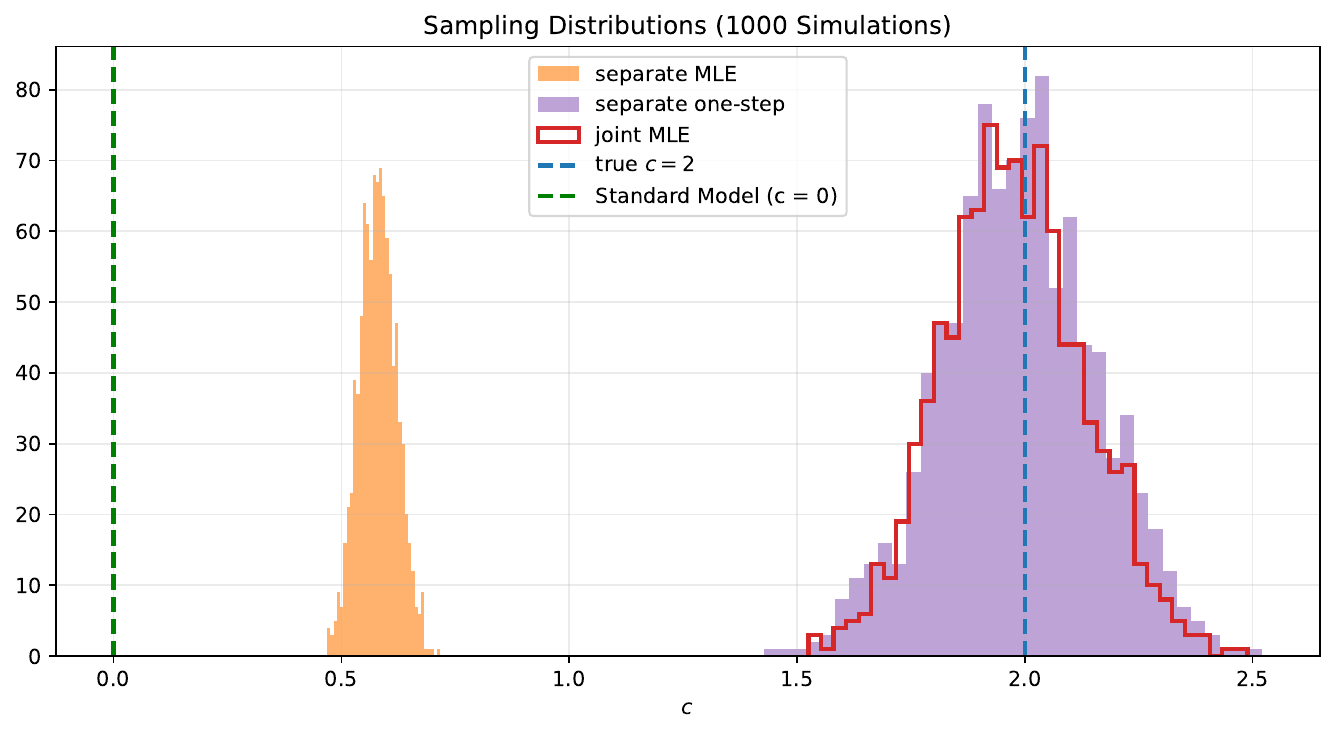}
    \caption{Sampling distributions of point estimates of the Wilson coefficient $c$. The data-generation and estimation procedure is repeated 1000 times. The filled orange histogram shows the sampling distribution of the preliminary separate MLE $\hat{c}_0$, the purple histogram shows the sampling distribution of the one-step estimator $\hat{c}_{\rm one\text{-}step}$, and the red outlined histogram shows the sampling distribution of the joint MLE of $c$. The vertical blue dashed line indicates the true value $c=2$, while the green dashed line indicates the Standard Model value $c=0$.}
    \label{fig:HERA_sampling_dist_c}
\end{figure}

\begin{figure}[htbp]
    \centering
    \includegraphics[width=\textwidth]{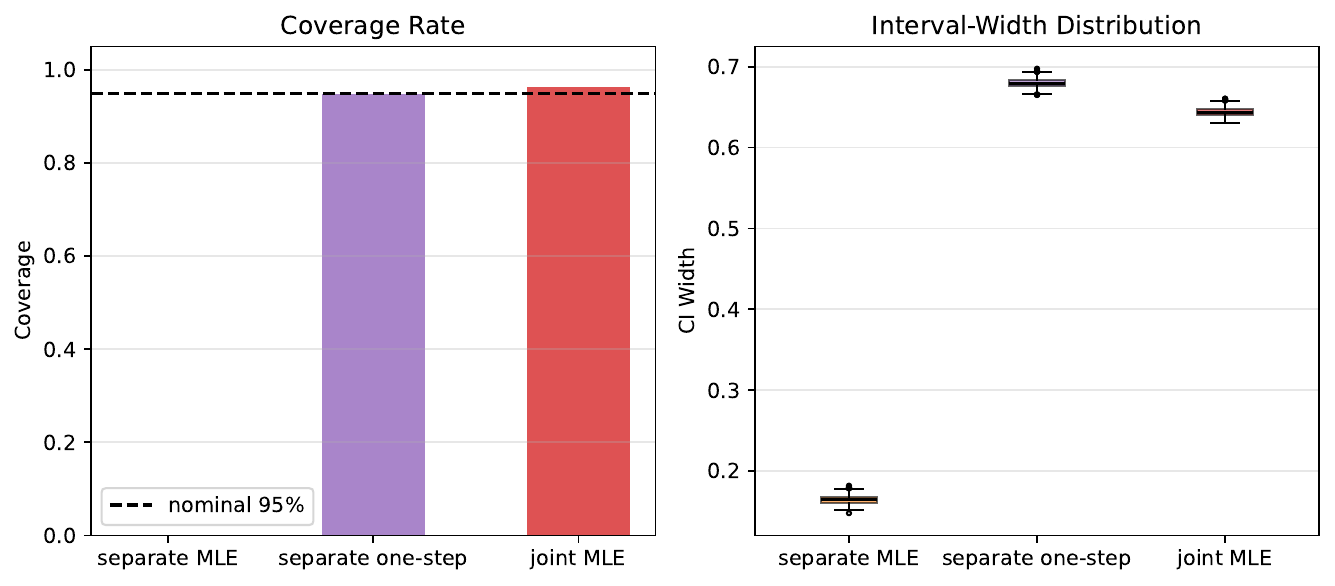}
    \caption{Coverage rates and confidence-interval widths for the Wilson coefficient $c$. The coverage rate is estimated as the proportion of the 1000 simulated confidence intervals that contain the true value $c=2$, with the black dashed line indicating the nominal 95\% coverage level. The right panel shows the distributions of confidence-interval widths, computed as the difference between the upper and lower interval bounds. Results are shown for the separate MLE $\hat{c}_0$, the one-step estimator $\hat{c}_{\rm one\text{-}step}$, and the joint MLE of $c$.}
    \label{fig:HERA_CI_coverage_width}
\end{figure}

\subsection{Results}
\begin{figure}[htbp]
    \centering
    \includegraphics[width=.8\textwidth]{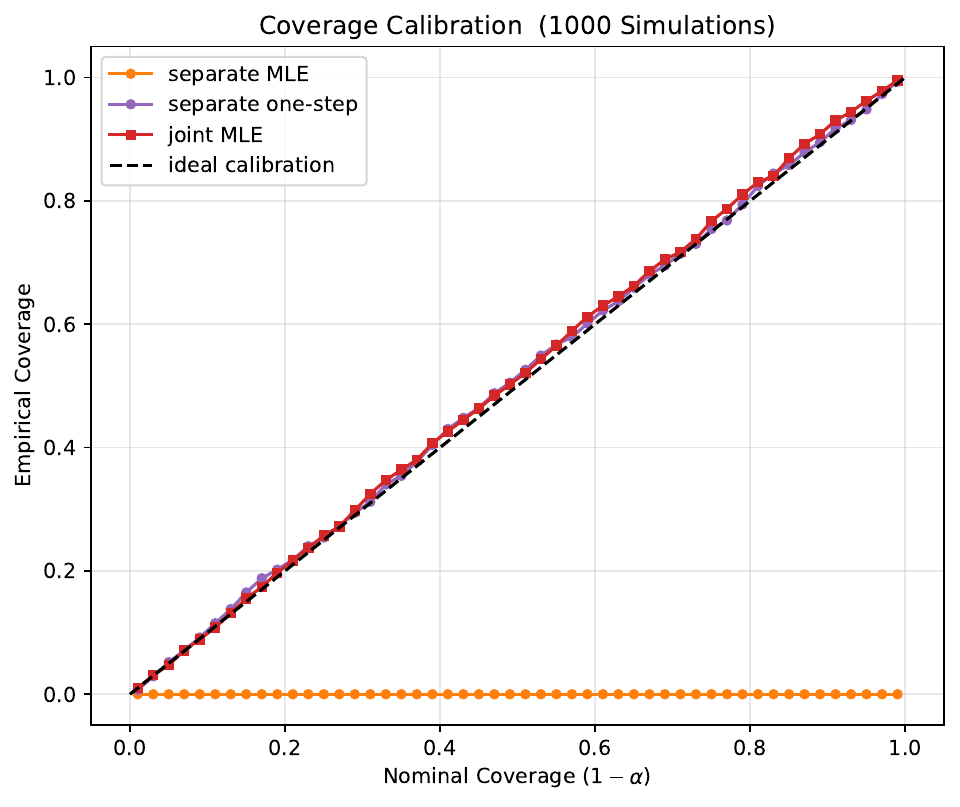}
    \caption{Coverage calibration of the confidence intervals for the Wilson coefficient $c$. The horizontal axis represents the nominal confidence level, while the vertical axis shows the empirical coverage rate estimated from 1000 simulations. The orange curve with circular markers corresponds to confidence intervals based on the separate MLE $\hat{c}_0$, the purple curve corresponds to intervals based on the one-step estimator $\hat{c}_{\rm one\text{-}step}$, and the red curve with square markers corresponds to intervals based on the joint MLE of $c$. The black dashed diagonal line represents ideal calibration, for which the empirical coverage matches the nominal confidence level.}
    \label{fig:HERA_coverage_calibration}
\end{figure}

The results are similar to those obtained in Sect.~\ref{sec:application_BCDMS}. That is, the preliminary estimate $\hat{c}_0$ is biased and systematically underestimates the true value of $c=2$, while the one-step estimator $\hat{c}_{\rm one-step}$ successfully corrects for this bias and yields confidence intervals with correct coverage. The data-level fit for $F_3^{\gamma Z}$ of one realization of pseudo-data is shown in Fig.~\ref{fig:HERA_data-fit} and the corresponding 95\% confidence intervals for $c$ are shown in Fig.~\ref{fig:HERA_CI_c}. As we can see, the bias in $\hat{c}_0$ is much more pronounced compared with the BCDMS case. Nevertheless, even in this more challenging setting, the one-step correction effectively removes the bias. The sampling distributions of the estimates of $c$ obtained from the separate MLE, its one-step correction, and the joint MLE are shown in Fig.~\ref{fig:HERA_sampling_dist_c}. The coverage and width of the confidence intervals for $c$ based on the three estimation methods are summarized in Fig.~\ref{fig:HERA_CI_coverage_width}, while the coverage calibration of the confidence intervals is shown in Fig.~\ref{fig:HERA_coverage_calibration}.

Overall, these results demonstrate that the preliminary estimator can suffer from substantial bias because changes induced by the BSM contribution can be partially absorbed into the estimated PDF. The one-step estimator effectively corrects this bias without having to refit the PDF and produces confidence intervals with substantially improved coverage and calibration. This is crucial for making valid inference about possible BSM effects, since biased estimates or miscalibrated intervals could lead to misleading conclusions about the size or significance of new physics contributions. Importantly, the performance of the one-step estimator is also close to that of the joint MLE, both in terms of the location of the sampling distribution and the resulting uncertainty quantification. Thus, the one-step approach can recover much of the inferential benefit of joint estimation while avoiding the need to perform a full simultaneous fit over the parameter of interest and the nuisance PDF.

\section{Discussion and Conclusion}
\label{sec:conclusion}
In this work, we developed a semiparametric framework for the simultaneous estimation of a physics parameter $c$ and uncertain PDFs, and demonstrated its effectiveness as a proof of concept in a controlled idealised setting. The central ingredient is a one-step estimator that debiases the preliminary estimate of $c$ by adding a correction term constructed from the efficient influence function. This has the large practical benefit that estimates of physics parameters extracted using biased PDFs can be corrected without having to refit the PDFs, so that a single
PDF can be reused across many possible SM parameters and BSM scenarios.

We first applied the proposed method to the case of BSM pseudo-data from the BCDMS experiment sensitive to the $T_3$ triplet PDF, followed by the extraction of the valence distribution $V$ from BSM pseudo-data of $F_3^{\gamma Z}$ measured at HERA. The results demonstrate that the one-step estimator effectively corrects the bias in the preliminary estimate and produces confidence intervals with correct coverage. Working with pseudo-data and known ground truth allows us to validate the methodology rigorously and establish a firm basis for applications to real experimental data.

It is worth noting that, in the present study, the joint MLE already provides an accurate estimate of $c$, as the PDF is represented by only eight basis functions, resulting in a relatively low-dimensional nuisance model. As the PDF model becomes more flexible and higher-dimensional, however, nuisance estimation becomes increasingly challenging, and the resulting estimation errors may induce non-negligible bias in the estimate of $c$. In such settings, the one-step correction is particularly valuable, as it can reduce the first-order effect of nuisance estimation errors on the target-parameter estimate. Although the correction is applied to preliminary estimates obtained from decoupled fits of $c$ and $f$ in the present study, it could also be extended to simultaneous fits in future work, if more sophisticated machinery for simultaneous fits becomes available.

More generally, the one-step correction is not tied to a particular method for constructing the preliminary estimate. The key requirement is that the data-level fit of the preliminary estimate should be consistent and sufficiently close to the data. When the preliminary estimate is already accurate, the empirical efficient influence function will be close to zero, and the one-step correction therefore has only a negligible effect. By contrast, when the preliminary estimate is biased due to nuisance PDF estimation error, the one-step correction can reduce this bias and thereby improve both estimation accuracy and the validity of subsequent inference.

Finally, this work represents a proof of concept demonstrating the promise of the semiparametric approach to the joint determination of PDFs and physics parameters. A key strength of the framework is its complete generality: it is not restricted to Wilson coefficients of an EFT, but applies equally to any physics parameter that enters the forward map as a smooth functional. From this perspective, SM precision parameters, such as the strong coupling $\alpha_s(M_Z)$, the top mass $m_t$, the $W$-boson mass $M_W$, or the weak mixing angle $\sin^2\theta_W$, are perhaps the most compelling next application, given the central role that PDF uncertainties play in limiting the precision of such determinations. Future directions also include extending the methodology to more PDF flavours and to more complex BSM scenarios involving multiple Wilson coefficients or more general EFT expansions. A further direction is the application to hadronic processes, where the cross section is bilinear in two initial-state PDFs rather than linear in a single one as in the DIS case studied here.

\begin{center}
\rule{0.6\linewidth}{0.4pt}
\end{center}

\noindent
The results presented in this work can be reproduced using the code available at the following repository: 
\begin{center}
\url{https://github.com/richardzhs/Semiparametric_Parameter_Estimation_PDF}
\end{center}

\acknowledgments

We are grateful to the PHYSTAT community for inspiring discussions during the early stages of this work. We especially thank Nina Elmer for providing assistance with setting up the toy model in Sect.~\ref{sec:application_BCDMS}. J.t.H and M.U would like to thank the members of the NNPDF collaboration for insightful discussions. The work of J.t.H is supported by the Science and Technology Facilities Council (STFC) via grant awards ST/T000600/1 and ST/X000494/1.  MU's work has been supported by the European Research Council under the European Union’s Horizon
2020 research and innovation Programme (PBSP, Grant agreement n.950246) and is partially supported by the STFC consolidated grants ST/X000664/1 and UKRI4224. 

\appendix
\section{Validity of the One-Step Estimator}
In this section, we establish the asymptotic validity of the proposed one-step estimator. We also discuss the interpretation and practical implications of the assumptions.
\label{sec:one-step-validity}
\begin{theorem}
Assume:
\begin{itemize}
    \item[(A1)] $\frac{1}{n}\|\delta\|_{\Sigma^{-1}}^2 \rightarrow \tau^2$, where $\delta = \dot{K}_cf - K_ch^*$,
    \item[(A2)] $\varphi(\bm{y};\hat{c}_0,\hat{f}_0) - \varphi(\bm{y};c,f) - \mathbb{E}[\varphi(\bm{y};\hat{c}_0,\hat{f}_0) - \varphi(\bm{y};c,f)] = o_p(n^{-1/2})$,
    \item[(A3)] $|\hat{c}_0-c|^2+\|K_{\hat{c}_0}(\hat{f}_0-f)\|^2 = o_p(n^{-1/2})$.
\end{itemize}
Then
\[
    \hat{c}_{\rm one-step} - c = \varphi(\bm{y}; f, c) + o_p(n^{-1/2}),
\]
and consequently
\[
    \sqrt{n}\left(\hat{c}_{\rm one-step} - c\right)
    \rightsquigarrow
    \mathcal{N}\!\left(0,\;\sigma^2    \right)
\]
where $\sigma^2 = 1/\tau^2$.
\end{theorem}
\begin{remark}
For Assumption (A1), the quantity
\begin{equation}
\frac{1}{n}\|\delta\|_{\Sigma^{-1}}^2
\end{equation}
represents the efficient information for c, which is normalized by the number of measurements. Thus, Assumption (A1) requires that this normalized information converges to a positive finite limit. This condition can be justified through more primitive assumptions on the residual efficient direction and the covariance structure. Specifically, let $\delta_i$ be the $i$-th component of $\delta$. Then
\begin{equation}
\frac{1}{n}\|\delta\|_{\Sigma^{-1}}^2
=
\frac{1}{n}\sum_{i=1}^n \delta_i^2 \Sigma^{-1}_{i,i}
+
\frac{1}{n}\sum_{i\neq j} \delta_i \delta_j \Sigma^{-1}_{i,j}.
\end{equation}
Therefore, a sufficient set of conditions for Assumption (A1) is as follows. First, suppose that the components of $\delta$ are uniformly bounded. Second, assume that the off-diagonal mass of the precision matrix is asymptotically negligible:
\begin{equation}
\sum_{i\neq j} |\Sigma^{-1}_{i,j}| = o(n).
\end{equation}
Under these two conditions, the off-diagonal term satisfies
\begin{equation}
\frac{1}{n}\sum_{i\neq j} \delta_i \delta_j \Sigma^{-1}_{i,j} = o(1).
\end{equation}
Thus, the normalized efficient information is asymptotically determined by its diagonal contribution. If, in addition,
\begin{equation}
\frac{1}{n}\sum_{i=1}^n \delta_i^2 \Sigma^{-1}_{i,i} \longrightarrow \tau^2 \in (0,\infty),
\end{equation}
then
\begin{equation}
\frac{1}{n}\|\delta\|_{\Sigma^{-1}}^2 \longrightarrow \tau^2.
\end{equation}
In other words, Assumption (A1) holds when the dependence between different measurement points is sufficiently weak, and the average diagonal contribution to the efficient information remains dominant and converges to a positive finite limit.
\end{remark}

\begin{remark}
    Assumption (A2) is an empirical process condition that requires the 
    difference between the estimated influence function and the true influence function, after centering by its expectation, to be asymptotically negligible at the $n^{-1/2}$ rate. This condition is a common requirement in semiparametric estimation, which requires the class of functions $\{\varphi(\cdot; c', f'): c' \in \mathbb{R}, f' \in \mathcal{F}\}$ belongs to Donsker class, which intuitively means that the class of functions is not too complex.
\end{remark}

\begin{remark}
    For Assumption (A3), although the statement requires $|\hat{c}_0-c|^2$ to be $o_p(n^{-1/2})$, in practice this might be relaxed. Specifically, if we have $\nu''(\hat{c}_0)=0$, then $|\hat{c}_0-c|^2$ can be dropped from the requirement. This can be satisfied for example if $\nu(c)$ is linear in $c$, i.e. BSM effect depends linearly on the Wilson coefficient. Even when $\nu(c)$ is not exactly linear—such as in our setting, where $\nu(c)$ is quadratic in $c$—the approximation can still perform well in practice, provided that the linear term dominates the contribution. Therefore, the main requirement for the preliminary estimator is $K_{\hat{c}_0}(\hat{f}_0-f) = o_p(n^{-1/2})$. In other words, the preliminary estimator can be relatively inaccurate in estimating the nuisance function $f$, as long as the resulting data-level fit is sufficiently close to the data.
\end{remark}

\begin{proof}
Let $P$ denote the distribution of $\bm{y}$ under the true parameters $(c, f)$, and $\hat{P}$ denote the distribution under the estimated parameters $(\hat{c}_0, \hat{f}_0)$. Our target parameter $c$ can be viewed as a functional of the distribution $P$, i.e.\ $c = \psi(P)$.
By the von Mises expansion, the efficient influence function $\varphi$ satisfies
\begin{align}
    \nonumber\psi(\hat{P}) - \psi(P) &= \int \varphi(x; \hat{c}_0, \hat{f}_0)\,d(\hat{P} - P)(x) + R_2(\hat{P}, P) \\
    &= -\int \varphi(x; \hat{c}_0, \hat{f}_0)\,dP(x) + R_2(\hat{P}, P)
\end{align}
where the second line follows from the fact that $\int \varphi(x; \hat{c}_0, \hat{f}_0)\,d\hat{P}(x) = 0$. $R_2(\hat{P}, P)$ is a second-order remainder term which depends only on products or squares of differences between $\hat{P}$ and $P$. Having $c=\psi(P)$ and $\hat{c}_0 = \psi(\hat{P})$ as well as the definition of the one-step estimator, we have
\begin{align}
    \nonumber\hat{c}_{\mathrm{one-step}} - c &= \hat{c}_0 + \varphi(\bm{y};\hat{c}_0,\hat{f}_0) - c \\
    \nonumber&= \hat{c}_0 + \varphi(\bm{y};\hat{c}_0,\hat{f}_0) - \left(\hat{c}_0 + \int\varphi(x;\hat{c}_0,\hat{f}_0)dP_{c,f}(x) - R_2(\hat{P}, P)\right) \\
    \nonumber&= \varphi(\bm{y};\hat{c}_0,\hat{f}_0) - \int\varphi(x;\hat{c}_0,\hat{f}_0)dP_{c,f}(x) + R_2(\hat{P}, P) \\
    \nonumber&= \varphi(\bm{y};c,f) + \varphi(\bm{y};\hat{c}_0,\hat{f}_0) - \varphi(\bm{y};c,f) - \int\varphi(x;\hat{c}_0,\hat{f}_0)dP_{c,f}(x) + R_2(\hat{P}, P) \\
    \nonumber&= \varphi(\bm{y};c,f) + \varphi(\bm{y};\hat{c}_0,\hat{f}_0) - \varphi(\bm{y};c,f) \notag\\
    \nonumber&\quad - \int(\varphi(x;\hat{c}_0,\hat{f}_0)-\varphi(x;c,f))dP_{c,f}(x) + R_2(\hat{P}, P) \\
    &= \varphi(\bm{y};c,f) + \mathbb{G}[\hat\varphi - \varphi] + R_2(\hat{P}, P)
\end{align}
where $\varphi = \varphi(\,\cdot\,; c, f)$ and $\hat\varphi = \varphi(\,\cdot\,; \hat{c}_0, \hat{f}_0)$, and
\begin{align}
    \mathbb{G}[\hat\varphi - \varphi] &= \hat{\varphi}-\varphi - \mathbb{E}_P[\hat{\varphi}-\varphi].
\end{align}
Notice that $\mathbb{E}_P[\varphi(\bm{y};c,f)] = 0$, and $\mathbb{G}[\hat\varphi - \varphi] = o_P(n^{-1/2})$ by Assumption (A2). Therefore, we have
\begin{equation}
    \mathbb{E}_P[\hat{c}_{\mathrm{one-step}} - c] = \mathbb{E}_P[R_2(\hat{P}, P)] + o(n^{-1/2}).
\end{equation}
For the remainder term, it is given by
\begin{align}
    \nonumber R_2(\hat{P}, P) &= \psi(\hat{P}) - \psi(P) + \mathbb{E}_P[\varphi(\bm{y}; \hat{c}_0, \hat{f}_0)] \\
    &= \hat{c}_0 - c + \mathbb{E}_P\left[\frac{(\dot{K}_{\hat{c}_0}\hat{f}_0 - K_{\hat{c}_0}h^*)^\top \Sigma^{-1}(\bm{y} - K_{\hat{c}_0}\hat{f}_0)}
           {(\dot{K}_{\hat{c}_0}\hat{f}_0 - K_{\hat{c}_0}h^*)^\top \Sigma^{-1}(\dot{K}_{\hat{c}_0}\hat{f}_0 - K_{\hat{c}_0}h^*)}\right].
\end{align}
Define $\mu(c,f)=K_cf=\nu(c)\odot \mathbf{T}_{\rm SM}(f)$ and write
\begin{equation}
    \mu = \mu(c,f), \qquad \hat{\mu} = \mu(\hat{c}_0, \hat{f}_0).
\end{equation}
Also, denote
\begin{equation}
    \hat{\delta} = \dot{K}_{\hat{c}_0}\hat{f}_0 - K_{\hat{c}_0}h^*.
\end{equation}
Then the remainder term can be written as
\begin{align}
    \label{eq:R2}
    R_2(\hat{P}, P) &= \hat{c}_0 - c + \frac{\hat{\delta}^\top\Sigma^{-1}(\mu-\hat{\mu})}{\hat{\delta}^\top \Sigma^{-1}\hat{\delta}}.
\end{align}
Taylor expanding $\mu$ around $(\hat{c}_0, \hat{f}_0)$ gives
\begin{equation}
    \label{eq:mu_expansion}
    \begin{split}
    \mu - \hat{\mu} &= (c-\hat{c}_0)\dot{K}_{\hat{c}_0}\hat{f}_0 + K_{\hat{c}_0}(f-\hat{f}_0) + \frac{1}{2}(c-\hat{c}_0)^2\ddot{K}_{\hat{c}_0}\hat{f}_0 + (c-\hat{c}_0)\dot{K}_{\hat{c}_0}(f-\hat{f}_0) \\
    &\quad +
    o_P(\|K_{\hat{c}_0}(\hat{f}_0-f)\|^2 + |\hat{c}_0-c|^2).
    \end{split}
\end{equation}
Notice that
\begin{align}
\nonumber\hat{\delta}^\top\Sigma^{-1}\dot{K}_{\hat{c}_0}\hat{f}_0 &= \hat{\delta}^\top\Sigma^{-1}(\dot{K}_{\hat{c}_0}\hat{f}_0 - K_{\hat{c}_0}h^*) + \hat{\delta}^\top\Sigma^{-1}K_{\hat{c}_0}h^* \\
    &= \hat{\delta}^\top \Sigma^{-1}\hat{\delta} + 0 = \hat{\delta}^\top \Sigma^{-1}\hat{\delta},
\end{align}
where the second equality follows from the definition of $\hat{\delta}$ and the fact that $K_{\hat{c}_0}h^*$ lies in the nuisance tangent space. Similarly,
\begin{align}
    \hat{\delta}^\top\Sigma^{-1}K_{\hat{c}_0}(f-\hat{f}_0) &= 0
\end{align}
since $K_{\hat{c}_0}(f-\hat{f}_0)$ lies in the nuisance tangent space. Therefore, the first two terms in the expansion \eqref{eq:mu_expansion} cancel out with $\hat{c}_0-c$ in the expression for $R_2(\hat{P}, P)$ in Eq. \eqref{eq:R2}, and we are left with
\begin{equation}
    R_2(\hat{P}, P) = \frac{1}{2}(c-\hat{c}_0)^2\ddot{K}_{\hat{c}_0}\hat{f}_0 + (c-\hat{c}_0)\dot{K}_{\hat{c}_0}(f-\hat{f}_0) +
    o_P(\|K_{\hat{c}_0}(\hat{f}_0-f)\|^2 + |\hat{c}_0-c|^2).
\end{equation}
Also, since
\begin{align}
    \ddot{K}_{\hat{c}_0}\hat{f}_0 &= \bm{\nu}''(\hat{c}_0) \odot \mathbf{T}_{\rm SM}(\hat{f}_0),
\end{align}
if $\nu''(\hat{c}_0)=0$, then the reminder term reduces to
\begin{equation}
    R_2(\hat{P}, P) = (c-\hat{c}_0)\dot{K}_{\hat{c}_0}(f-\hat{f}_0) + o_P(\|K_{\hat{c}_0}(\hat{f}_0-f)\|^2).
\end{equation}
Therefore, we conclude that $R_2(\hat{P}, P) = o_P(n^{-1/2})$ under Assumption (A3). This implies that $\mathbb{E}_P[\hat{c}_{\mathrm{one-step}} - c] = o(n^{-1/2})$. Finally, since
\begin{equation}
    \mathrm{Var}_P\{\varphi(\bm{y};c,f)\}
    =
    \frac{1}{\delta^\top \Sigma^{-1}\delta},
\end{equation}
Assumption (A1) implies
\begin{equation}
    n\,\mathrm{Var}_P\{\varphi(\bm{y};c,f)\}
    \rightarrow \frac{1}{\tau^2}.
\end{equation}
Therefore,
\begin{equation}
    \sqrt{n}(\hat{c}_{\mathrm{one-step}} - c)
    =
    \sqrt{n}\varphi(\bm{y};c,f) + o_P(1)
    \rightsquigarrow
    \mathcal{N}\!\left(0,\frac{1}{\tau^2}\right).
\end{equation}
\end{proof}

\bibliographystyle{JHEP}
\bibliography{references}

\end{document}